# High Strain Engineering of a Suspended WSSe Monolayer Membrane by Indentation and Measured by Tip-enhanced Photoluminescence.

Anis Chiout [1], Agnès Tempez [2], Thomas Carlier [2], Marc Chaigneau [2], Fabian Cadiz[3], Alistair Rowe [3], Biyuan Zheng [4], Anlian Pan [4], Marco Pala [1], Fabrice Oehler [1], Abdelkarim Ouerghi [1], Julien Chaste [1*]

[1] Université Paris-Saclay, CNRS, Centre de Nanosciences et de Nanotechnologies, 91120, Palaiseau, France.
[2] HORIBA France SAS, Passage Jobin Yvon, 91120 Palaiseau, France
[3] Laboratoire de Physique de la Matière Condensée, CNRS, Ecole Polytechnique, Institut Polytechnique de Paris, 91120 Palaiseau, France.
[4] Key Laboratory for Micro-Nano Physics and Technology of Hunan Province, State Key Laboratory of Chemo/Biosensing and Chemometrics, and College of Materials Science and Engineering, Hunan University, 410082 Changsha, Hunan, China.

* Corresponding author: julien.chaste@universite-paris-saclay.fr

**Abstract**

Straintronics involves the manipulation and regulation of the electronic characteristics of 2D materials through the use of macro- and nano-scale strain engineering. In this study, we utilized an atomic force microscope (AFM) coupled with an optical system to perform indentation measurements and tip-enhanced photoluminescence (TEPL), allowing us to extract the local optical response of a suspended monolayer membrane of ternary WSSe at various levels of deformation, up to strains of 10%. The photoluminescence signal is modelled considering the deformation, stress distribution and strain dependence of the WSSe band structure. We observe an additional TEPL signal that exhibits significant variation under strain, with 64 meV per percent of elongation. This peak is linked to the highly strained 2D material lying right underneath the tip. We discuss the amplification of the signal and its relation to the excitonic funnelling effect in a more comprehensive model. We will also compare the diffusion caused by Auger recombination against the radiative excitonic decay. We use TEPL to examine and comprehend the local physics of 2D semi-conducting materials subjected to extreme mechanical strain. Chemical vapour deposition-fabricated 2D ternaries possess high strain resistance, comparable to the benchmark $MoS_2$, and a high Young's modulus of 273 GPa.

The controlled application of significant strain to two-dimensional (2D) materials, ranging from one to a few monolayers in thickness, has unveiled interesting physical phenomena and is commonly referred to as 2D straintronics. [1–3] This technique has potential in tuning specific absorbance or photoluminescence (PL) lines, [4–7] as well as producing quantum emission sources at near-room temperature. [8–12]

This fruitful combination arises from the unique physics of monolayer 2D transition metal dichalcogenides (TMDs). The majority of TMDs ($MoS_2$, $WSe_2$, $WS_2$, WSSe) are direct band gap semiconductors that exhibit complex quasi-particle physics dominated by excitonic effects, even at room temperature. [13] Combined with the innate resilience of two-dimensional materials to high strains (up to 6-11% in $MoS_2$ [14], 7% for $WSe_2$[15], 3% for $WS_2$[16]), monolayer TMDs are an ideal testing ground for observing the impact of extreme strain on electronic band structure. At room temperature, [17] excitonic peaks have demonstrated a remarkable tunability of the emission wavelength for direct band gap semiconductors that operate within the visible range, with lines shifting reversibly by several hundreds of meV. These findings convey noteworthy results concerning the effect of extreme strain upon electronic properties.

In these 2D materials, the physics is uniquely influenced by the combination of strain and excitonic quasi-particles. It creates a funnelling effect that leads to controlled spatial drift of excitons over several microns.[18–21] This phenomenon has facilitated recent experiments on exciton scattering at low temperatures,[22–24] where strain replaces the electric field and manipulates both charged and neutral excitons.[16,25] Several studies have been conducted on this topic. In particular, the dark excitons found in TMD materials are most sensitive to straintronics effect, since they exhibit a much slower decay rate than bright excitons, leading to longer diffusion times (and lengths) along the strain-induced energy gradient.[26,27]

The impact of mechanical strain on 2D TMD materials has been thoroughly examined, covering area such as crystalline anisotropy,[28] van der Waals inter-layers coupling control,[29] and even crystalline phase transition triggering.[30,31] As a result, the community has developed various approaches to strongly deform 2D materials, such as support-induced mechanical mismatch,[18,19,32] support bending,[4,28] substrate nanostructuring[5,33,34] [8,35,36] and comb-drive actuation.[37–39] Alternatively, 2D materials can also be deformed trough lattice displacement[40,41], angular misorientation between separate layers, [42] direct nanostructuring,[43,44] thermal heating,[45] or nano-indentation using atomic force microscopy (AFM) tip[16,22–27]. For 2D materials studied in a suspended membrane configuration, nano indentation using an AFM tip is possibly the most straightforward technique for the application of a local strain in a controlled manner, from minute strains up to extreme deformations exceeding ten percent.[46]

Understanding how mechanically induced deformation affects the electronic properties of atomically thin layers requires a deformation method with tunable strain coupled to a probe with a nanometer-scale resolution. Both of these requirements can be met with an AFM-Raman-TEPL system capable of tip-enhanced optical spectroscopy measurements, which selectively excite the material located near the tip apex and collect the corresponding signal. The movement of the sample, in combination with the amplification of the photoluminescence signal at the tip's apex in the near field, produces a tip-enhanced PL (TEPL) image. This offers spatial resolution of up to 20 nm [25,47,48]. Previously, [47] this AFM-Raman-TEPL system

(HORIBA) was used to observe the room temperature behavior of excitons in monolayer $WSe_2$ nanobubbles. This highlighted the localized exciton spatial zones that were tens of nanometers wide. Small, localized deformations of up to 0.2% were observed when a tip was pressed onto nano-wrinkles in a 2D material.[48] Additionally, a 10 meV shift in the photoluminescence peak was detected at nano-resolution using the TEPL technique, demonstrating the efficacy of this approach for 2D materials submitted to strain. Bubble made of 2D materials are stable system compatible with precise measurements of strain and nanophotoluminescence.[49,50] It stays a simple and very interesting experimental approach but it lacks a deterministic localization and reproducibility of shape or strain. Indentation technic induced a strongly tunable strain in the 2D material at the tip localization. However, if extended to larger deformations exceeding 10%, further discussion is necessary for the existing model of deformation and band structure in a domain where the band energy shift is of the order of the band gap itself.

In the current investigation, a schematic diagram of the AFM probe is presented in Figure 1a . The probe has been utilized to apply a calibrated localized force on a suspended 2D TMD WSSe monolayer membrane, which resulted in a circular strain distribution with a reverse "tent-shaped" dip. Figure 1b depicts the expected change in the band structure of the WSSe monolayer, with a gradual decrease in the optical band gap of the WSSe monolayer due to the layer deformation ε. This deformation is peaked at the tip apex (r=0) and decreases as we approach the tip center (r=∞→0). This produces a circular potential well, directing the excitons (electron-hole pairs) towards the region with the lowest band gap energy (i.e. the tip apex).

**Experiment**
The 2D material consists of large flakes of monocrystalline single layer WSSe alloy. We transfer the 2D material onto a hole-patterned silicon surface to suspend and stretch significant areas of the 2D crystal in a flat and circular "drum" configuration. The cavities are fabricated using the standard lithography technique on silicon/$SiO_2$ wafers where 300 nm deep holes are created via reactive ion etching.[51] A thin, 200 nm-wide evacuation channel connects the circular cavities to one another and to the external atmosphere. It ensures that no differential pressure is induce upon the membrane.

The large triangular flakes of monolayer WSSe used in this study are made of monocristaline materials. They have a width exceeding 100 µm and were grown on a $SiO_2$ substrate using vapor phase growth at high temperature (1100°C) in a furnace under an argon atmosphere.[52] WSSe is a ternary 2D TMD material whose optical gap varies based on the chalcogen stoichiometry. The chemical structure of our membrane is $WS_{1.3}Se_{0.7}$, resulting in a main PL emission near 1.85 eV in ambient conditions, when unstrained. [52,53] The WSSe monocrystals are transferred one-by-one onto the patterned substrate trough the following procedure; (1) coat the growth substrate with poly(methyl methacrylate) (PMMA); (2) immerse the sample in liquid nitrogen to gently detach the 2D material from the $SiO_2$ due to thermal compression mismatch[54] and clean it in acetone and isopropanol; (3) pick up the 2D layer in a transfer station using a polydimethylsiloxane (PDMS) half-sphere covered with PMMA;[55] (4) A drop of water should be inserted between the substrate and the PDMS bubble before contact to overcome the remaining interaction forces between the 2D and the $SiO_2$ surface.[56] This is due to the additional capillary force provided by the water which allows for the removal of complete sheets without apparent damage; (5) The PDMS/PMMA/2D layer is then precisely aligned with the patterned structure; (6) The PMMA/2D assembly is transferred by slowly heating the PDMS to 180°C; (7) The PMMA is then completely annealed at 350°C under vacuum for several hours.

This fabrication procedure gives a flat, homogeneous, and optically defect-free single layer of monocrystalline WSSe fully covering the cavities. An optical image is shown in Figure 1c (top view) with the circular cavities appearing dark over the purple silica surface. The figures 1d and 1e show the corresponding AFM topography and AFM phase images of two adjacent "drums", respectively. The topography reveals a near flat surface while the phase image clearly separates the (mechanically soft) suspended membrane area from the membrane lying on the (relatively harder) silicon surface. In the absence of external strain, the room temperature photoluminescence of the suspended WSSe membrane consists in two standard peaks, separated by 40 meV, namely the A exciton peak at 1.872 eV and the trion peak A' at 1.832 eV, as shown in Figure 1f. The Raman response (not shown) and the absence of other PL contributions confirm the stoichiometric nature of our WSSe material and the absence of any surface oxidation.[53]

First, a TEPL cartography was conducted on a WSSe flake for the first time, using a silver-coated tip stimulated by a 532 nm laser with a power of P=30 µW. The results are presented in Figure 1g. To acquire the genuine nanoPL response of the material, the PL signal in the "far-field" (tip a few nanometers away from the surface) was subtracted from the "near-field+far-field" PL signal (tip in contact with the surface). In our measurement, TEPL typically corresponds to ~10-20% of the "far-field" signal. [57–61] We observe that the Au-coated tip, which create a metal-semiconductor contact does not affect the band structure of the WSSe (See part S10 of the supplementary information)

The determination of the spatial extension near the tip apex that is responsible for the TEPL emission is a crucial parameter for accurate TEPL signal modelling. An experimental evaluation of this region extension was performed and is presented in the Supplementary Information. The area was approximate by a Gaussian profile with a FWHM of $w_{TEPL}$ =90 nm. We carried out spatial calibration of the AFM tip and determined a tip diameter of $2r_{tip}$ = 40 nm from AFM topography, as well as an associated cantilever stiffness constant $k_{tip}$ = 3.11 N/m (refer to Supplementary Information). The tip diameter and stiffness align with the supplier's provided specifications.

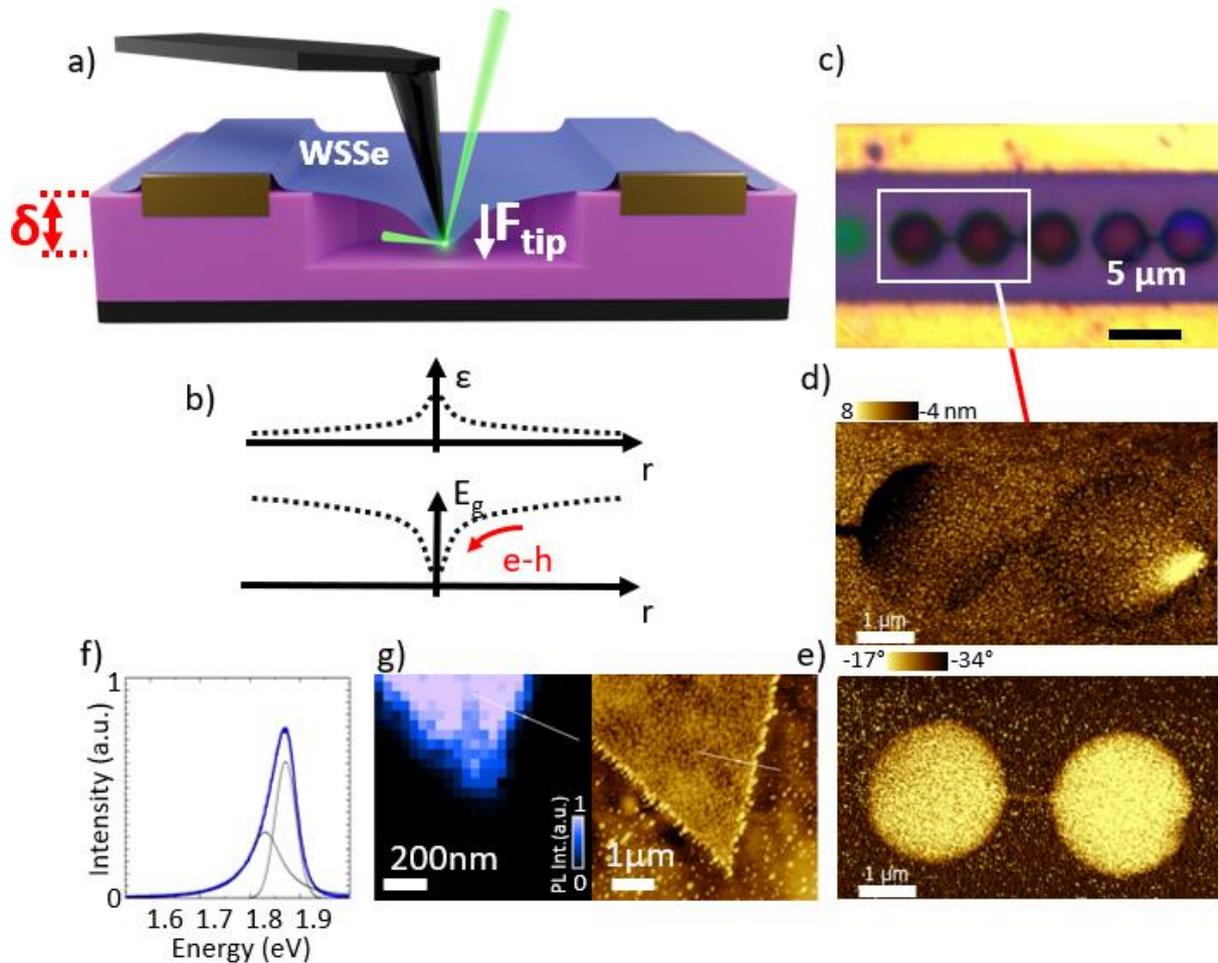

**Figure 1: Optical measurement of a suspended WSSe monolayer in an AFM tip indentation experiment.** a) The experiment is shown schematically, in which an AFM tip applies a mechanical force to the center (r = 0) of a suspended 2D monolayer membrane. b) Indentation alters the deformation at the center of the membrane and the local band gap. Due to scattering and drift, the excitons quickly move towards the centre and are subsequently re-emitted at the lowest energy level of the band gap. Additionally, there is an optical image of the sample provided, along with corresponding topography (d) and phase (e) images of the 2D drum. Notably, the suspended part is completely level. It is observable that there is a thin trench between the two drums. f) The photoluminescence spectrum of WSSe with no strain is displayed, with a two-peak fit. g) The TEPL signal of a 2D flake is measured along its edge, and the corresponding topography is measured using atomic force microscopy.

After calibrating the HORIBA AFM-Raman-TEPL, we can proceed with the indentation measurements using the same optical excitation (532 nm, 30 µW). Figure 2a displays the "near-field+far-field" PL, where the metal tip and laser spot are aligned. A noticeable broadening and shifting of the PL signal is observed as the indentation depth ($\delta$) increases, followed by the emergence of numerous additional peaks. We conducted the same indentation experiment again, but with a 300 nm laser spot offset, to obtain the "far-field" PL (Figure 2b). This distance ensures that the signal is free from tip enhancement and collects only the µPL or far-field signal. At first glance, the "far-field" PL exhibits similar behaviour to the "near-field+far-field" PL, broadening and red-shifting with increased indentation depth. However, they differ due to the "TEPL/nanoPL" signal.

The TEPL signal (Figure 2c), which is calculated by subtracting the "far-field" PL signal from the combined "near-field+far-field" PL signal, displays the most red-shifted contribution of the PL signal at each indentation depth. It is accompanied by a clear monotonic peak shift (from 1.87 eV to 1.74 eV) but is dominated by a single component, unlike the "near-field" and "far-field" data. Considering that TEPL mainly measures the PL response from the area directly beneath the tip apex, its variation reflects the significant modification of the optical response of the suspended WSSe monolayer under extreme tensile strain. In the following section, we present a model to simulate these optomechanical effects and confirm the nature of the TEPL signal.

Figure 2c illustrates the as-measured TEPL signals as a function of strain at the tip apex. The TEPL peak has intensified from 690 counts to 2350 counts for different strain. This indicates that the PL intensity has increased by at least 265 % in TEPL, when probed area is reduced by a factor of 36, and that we can isolate the TEPL signal. In our computations, we incorporate a far-field residual signal (540 counts) to fit the experimental data, as we do not entirely remove the tip in our experiment. In the SI, we examine the signal response at the indentation point and 300 nm away from it, as depicted in Figure 2. However, this correction alone is inadequate in explaining our data.

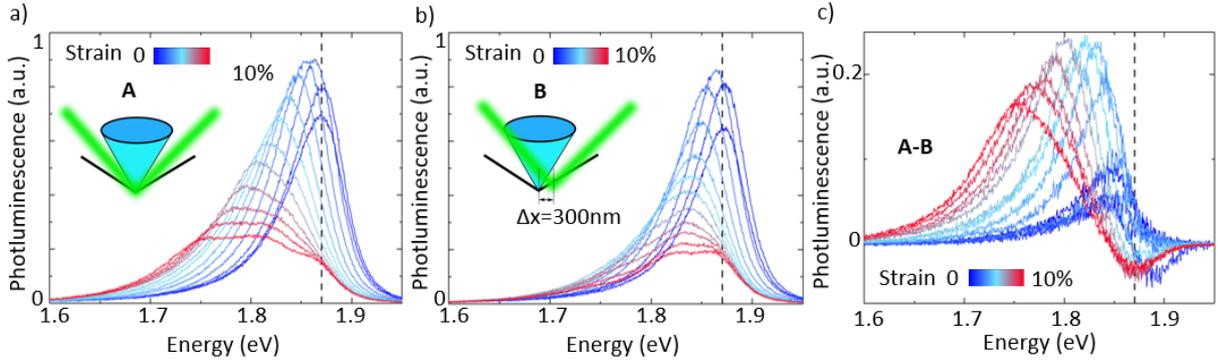

**Figure 2:** Operando photoluminescence measurements were obtained under mechanical pressure. Tip-enhanced photoluminescence spectra were collected with the laser directed towards the tip apex (TEPL "hot spot") for varying deformation depths. Far-field PL spectra were also collected with the laser spot positioned 300 nm away from the tip apex for different deformation depths. Pure near-field PL (true TEPL) spectra were generated by subtracting the far-field spectra "b" from spectra "a".

**Discussion**

To determine the optical response of the material at the membrane's contact point, we must consider the band gap and strain distribution.

The following three steps will be followed: 1) an estimation of the strain profile along the membrane, $\varepsilon(r \in [0, R])$, using radial coordinates, where R represents the radius (cavity radius) of the membrane, which is 1.25 µm. This step is necessary due to the collecting optics' spatial distribution and the tip's non-zero radius. We calculate parameters, including Young's modulus, by matching the force-deflection curve, $F(\delta)$, and obtaining the maximum strain $\varepsilon_{max} = \varepsilon(r = 0)$ at any depth of indentation. 2) Subsequently, using density functional theory (DFT), we establish the correlation between a particular local strain at any r and the local bandgap $E_G$. 3) Next, we convolve this spatial distribution of strain and bandgap with the local population of excitons to derive the standard photoluminescence (PL) signal from the sample.

To describe the strain $r \in [0, R]$, we must consider two areas: the central region of tip-membrane contact, with high strain, and the suspended region of the membrane, with lower strain. Two models will be used to describe the strain along the sample.

The first is a flat tip model, which employs a simple model of a flat tip. The Schwerin solution fixes the strain at r = 0. A point loading model, like the widely recognized Schwerin solution, gives a reliable estimation of the maximum strain induced by indentation. To expand this model to any radius, we presume a steady stress along the tip contact and apply an empirical model that fluctuates as $1/r$ in regions outside the tip. This approximation has been demonstrated as reasonable for indented 2D materials. [7,22]

The second model utilizes the Föppl–von-Kármán equations and Föppl-Hecky small finite-deflection theory. This model provides specific treatment of the contact region, resulting in an accurate representation of indentation. The Bathia and Nachbar indentation model, as we refer to it, or rounded tip model, is a more formal model featuring a rounded tip. [62–65]

In Figure 3a, we apply a force F to the center of the circular membrane that deforms elastically the 2D with a central deflection δ. Our experimental data (blue points) reaches δ=250 nm for F=2200 N (Figure 3a, log scale data). Such an indentation experiment was repeated several times, with identical results, demonstrating the elastic character of the deformation. We first consider a standard Schwerin indentation model,[66,14,67,15,68] which relates F to δ while approximating the tip by a point. Knowing F and δ, the Schwerin model $F = \sigma_{2D}.\pi\delta + E_{2D}.q_1^3/R^2 . \delta^3$ is used to evaluate the initial 2D stress $\sigma_{2D}$ and stretching modulus $E_{2D} = E.t$ of the WSSe membrane. Here, E is the bulk Young modulus, t=0.7 nm is the membrane thickness and the parameter $q_1$ is defined as $q_1 = 1/(1.05 - 0.15.\nu - 0.16\nu^2)$ with $\nu = 0.19$ the Poisson ratio of the bulk material for $WS_2$ and $WSe_2$.[69,70] From the fitting to the experimental results (the red line in Figure 3a) we obtain $\sigma_{2D} = 0.158$ N/m and $E = 272.7$ GPa for our system. Young's modulus of the ternary $WS_{1.3}Se_{0.7}$ is in accordance with previous measurements in $WS_2$ (E ~ 250 GPa) and in $WSe_2$ (E ~ 300 GPa). [71] We describe the strain at the apex $\varepsilon_{max} = \sqrt{F/4\pi.r_{tip}.Et}$ and we plot it as a function of the deflection in Figure 3b. In Figure 3b, the maximum strain, > 10 %, is obtained with a 250 nm deep indentation. This confirms that indentation is an appropriate technique to deform 2D material to their limit and a solution to explore the physics of 2D materials under very high strain.

In the rounded tip model, the tip has a radius of curvature $r_{tip2}$ and is in contact with the 2D material below a radius $r_{contact}$, and this contact radius increases with the indentation deflection δ. It gives a $1/\sqrt{r}$ dependence for the strain distribution far from the tip (see Figure 3c and SI for details). After calibration of the required parameters (see Supplementary Information), if we consider only the strain at r = 0, the Bathia and Nachbar model correctly describes our experimental F(δ) data and leads to a comparable Young's modulus and $\varepsilon_{max}$ (Figure 3b). To account for the experimental shape of the tip measured during the AFM measurement, which corresponds to very small indentations ($\varepsilon < 0.5\%$), we must have $r_{contact}=r_{tip}=20$ nm, which fixes $r_{tip2} = 60$ nm.

To know exactly the effect of the indentation on the optical response of the material, it is crucial to determine both the strain profile along the membrane and the effect of the strain on the band structure of the 2D material. In Figure 3d, calculations have been performed to evaluate the variation of the band structure of WSSe subjected to a biaxial strain of the order of a few percent. We considered the $WS_{1.3}Se_{0.7}$ using density functional theory (DFT). We have adopted the hybrid Heyd-Scuseria-Ernzerhof (HSE) function approximation as well as the virtual crystal approximation. The coupling of spin orbitals is included in our model. The cell and atomic positions of the undeformed WSSe were evaluated after full relaxation with cell parameters $a_0$=3.24 Å and c=28 Å. A bi-axial in-plane strain was imposed on the unit cell and then the atoms were allowed to relax their positions. As seen in Figure 3e, the resulting gap at K point decreases (increases) under a biaxial tensile (compressive) stress. For large compressive strain ($\varepsilon_x < -1\%$), the conduction band minimum moves from the K point to the Q point and the associated optical transition becomes indirect. For tensile strains, as in our indentation experiment, the material keeps its direct band gap character with increasingly smaller value. To first order, we model this variation as $E_G = E_{G0} - dE_G/d\varepsilon_x \cdot \varepsilon_x$, with $dE_G/d\varepsilon_x = $ 125meV/%. To link the DFT calculation using biaxial deformation ($\varepsilon = \varepsilon_x = \varepsilon_y$) to our radial strain model, we define $\varepsilon_r^2 = \varepsilon_x^2 + \varepsilon_y^2$ and we obtain $dE_G/d\varepsilon_r = 1/\sqrt{2} \cdot dE_G/d\varepsilon = $ 88meV/%. In this way we can now model the local band gap value at the radial elongation r from the tip as $E_G = E_{G0} - dE_G/d\varepsilon_r \cdot \varepsilon(r)$,

To confirm the theory, we also performed photoluminescence and optomechanical measurements of the static strain on the 2D membrane with different gate voltages $V_g$. The gate voltage is used to pull-in the entire membrane and to scan different deformations. The suspended membrane vibration is measured by electrically activating the out-of-plane vibration and measuring it by optical reflectometry.[30,51] The mechanical frequency versus gate voltage dependence can be used to extract the strain along the membrane (see SI). We extract the optical response as a function of the strain applied to the membrane and observe a 5 meV shift in the peak position for 0.04% of elongation, i.e. 125 meV/% along x which corresponds quite well with the theory. Also, these measurements confirm the good mechanical quality of the membrane with high quality factor measurement Q=1000).

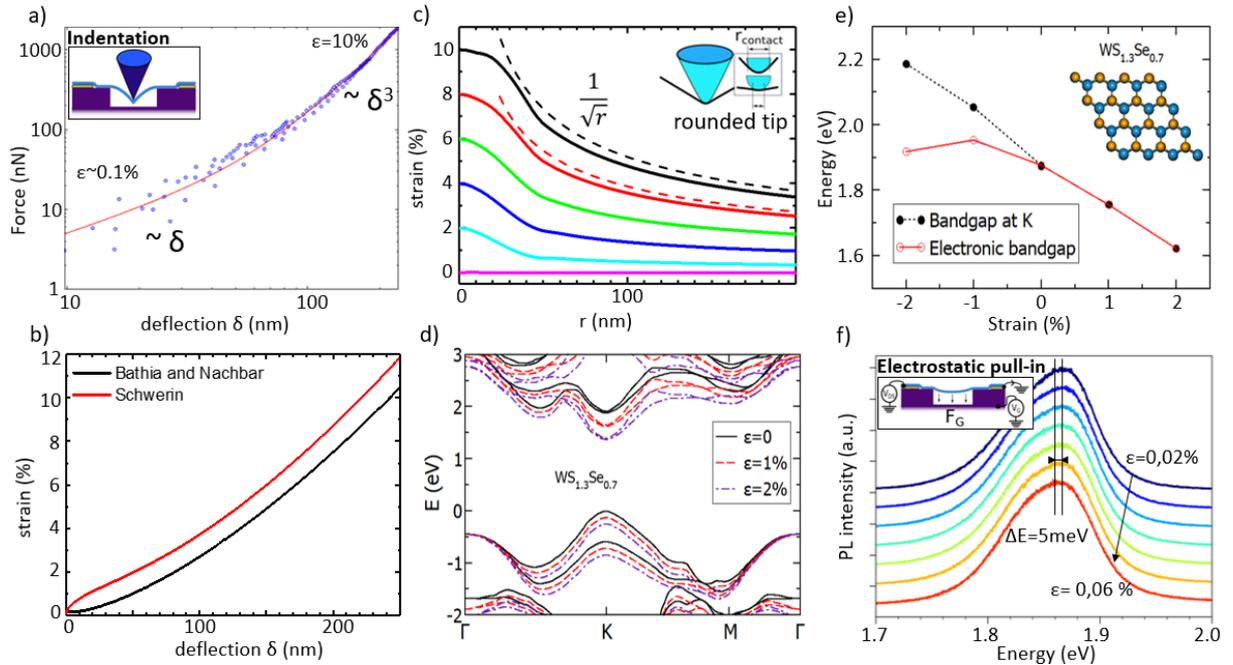

**Figure 3: Indentation induced local strain, band gap tuning and exciton/charge funnelling.**
a) A force indentation experiment with the force applied to the membrane as a function of the membrane deflection. Two regimes are distinguished at low and high deformations. b) The strain at the r=0 as a function of the deflection for the Schwerin model [68,14,67,15,66] and the Bathia and Nachbar model.[62,63] c) The strain as a function of the radius for different force values with the analytical model of a round tip. 10% strain corresponds to F~2200 nN. d) Band structure of the WSSe calculated for different biaxial elongation applied to both x and y directions. The electronic structure varies strongly at the K point. e) Based on the calculations shown in d), we plot the band gap, $E_G$, as a function of the strain applied along x and y ($\varepsilon = \varepsilon_x = \varepsilon_y$). e), f) A global deformation of the membrane is applied by a capacitive force, which induces a small PL peak shift. The corresponding strain is extracted from optomechanical measurements.

The µphotoluminescence response is the convolution of the local PL spectra at the position excited by the laser spot and the local density of excitons, n(r). In a first approximation, without excitonic drift or funnelling, n(r) is thus directly proportional to the Gaussian intensity profile of the laser beam, $w_{Laser} = 600$ nm. For the TEPL measurements, the width of the Gaussian distribution is reduced to $w_{TEPL}$. Under these assumptions we define the simulated ⟨PL⟩ signal as $\langle PL \rangle = \int PL(E_G(r))n(r)rdr$. The simulated and experimental PL response are plotted as a function of the strain applied to the membrane center, up to 10%, for the far-field signal in Figures 4a, b, c, and for the TEPL signal in Figure 4e, d, f. We compare the experimental results with the simulations for the two types of strain distributions detailed earlier, the one with a flat tip and a $1/r$ dependence of the radial strain, and the second with a rounded tip and a $1/\sqrt{r}$ dependence of the radial strain. For small strains ($\varepsilon < 3\%$) the experimentally measured PL peak does not shift much, as is the case for both tip models. This is probably related to the fact that the contact remains point-like at this stage, with the area of contact between the tip and the 2D stay closed to the value measured by AFM, and that the stress area is much smaller than the spatial extension of the laser. In this region, the flat tip with constant radius is a good approximation of the system. In the limit of strong deformations (>5%), the signal shifts more in the case of a rounded tip (25 meV/%) than for a flat tip (2.5 meV/%). The exciton peak shift (black curve) for the measurement is 22 meV per % of elongation for a strain >5% in the Figure 4b. In this high strain region, it seems appropriate to describe the AFM probe as a rounded tip with a $1/\sqrt{r}$ dependance, where the contact radius changes with strain and the 2D material surrounds the tip.

In Figure 4b, the evolution of the µPL is depicted as a function of the strain applied to the sample. The peaks were analyzed using Gaussian or Lorentzian fits ranging from 2 to 5. It is noteworthy that the peaks are consistently separated by 55 meV. We observe a correlation between this energy and the trion binding energy, but it is not directly related to quantization of the excitonic levels in a low-dimensional system,[72] which are expected at much lower energy (~a few meV) for systems of a size equivalent to the area with a high strain gradient (~50 nm).

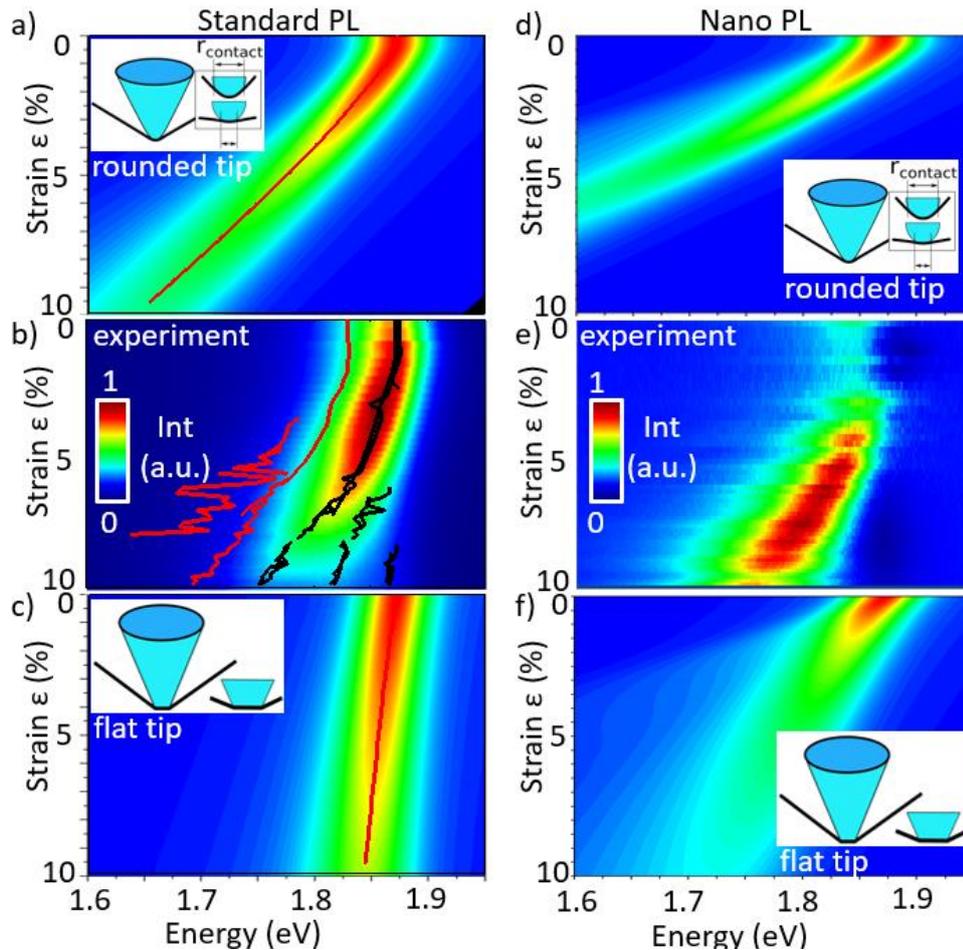

**Figure 4 Model and experiment** a) and c) The far-field PL response is compared as a function of the strain, simulated for the rounded tip and the flat tip respectively, with the measured PL response in b). b) The positions of the peaks obtained for a two-, three- or four-peak fit for each strain value. We distinguish in black the peak that seems to be linked to the A exciton and in red the peaks that seem to be associated to the A' trion. d) and f) The TEPL responses as a function of the strain, simulated for the rounded tip and the flat tip respectively, are compared with the measured TEPL response in e).

Finally, now that we have properly calibrated the indentation and globally understood the PL response of the system, we focus our attention on the TEPL signal in Figure 5a and 5b. In particular, we address the existence of additional shoulders at lower energy, between 1.68 eV and 1.8 eV and at "lower" strain, from 3 to 5%, which continuously appeared. The amplitude of this peak remains very weak compared to the main PL peak signal intensity, about a factor of 8 lower in intensity, but nevertheless detectable over several indentation measurements (the different symbols in Figure 5-c). This peak moves in the corresponding strain and energy windows with a slope of 64 meV/%. It is quantitively accurate to relate this peak to the PL response at the maximum strain position (88 meV/%), just below the tip for a strain between 3 and 5%. It is therefore possible to measure the photoluminescence signal coming from a nano-area of a 2D material, located around an AFM tip, and strongly stressed, from 3 to 5%, during an indentation experiment at room temperature.

We will now discuss the importance of the funnelling effect in the appearance of this shoulder peak. As predicted earlier,[73] the scattering and the mobility of carriers tend to move the electron-hole pairs to the area of lowest band gap, i.e. greater elongation area, before the radiative emission of excitons or the Auger recombination. It must enhance the peak amplitude of the TEPL signal. It is necessary in this case to redefine the density of excitons n(r) and the TEPL signal by considering the funnelling, as in Figure 5d. In our simulations, in Figure 5e, it is possible to recover a signal with a slope of 82 meV/% which corresponds to the band gap variation of the 2D material under the tip, at maximum strain. At this stage, in order to obtain the real hyperspectral photoluminescence image of our sample, we still have to determine the efficiency of the funnelling effect, the exciton drift and diffusion along the sample, submitted to the band gap $E_G(r)$. The density of excitons n(r) must follow the drift-diffusion equation.

$$\nabla\big(D\nabla n(r)\big) + \nabla\big(\mu\, n(r)\nabla E_G(r)\big) - n(r)/\tau - n^2(r)R_A + S(r) = 0$$

S(r) is the exciton generation function with the gaussian distribution of the laser beam, $S(r) = I_0/2\pi\sigma^2 e^{-r^2/2\sigma^2}$. With w the half width laser size (600 nm), $\sigma = w/2\sqrt{2\ln(2)}$ and $I_0$ the laser intensity. $1/\tau = 0.91\ \text{ns}^{-1}$ and $R_A = 1,4.\,10^{-5}\,\text{m}^2.\,\text{s}^{-1}$ are respectively the exciton decay rate and the Auger recombination rate. D is the diffusion coefficient and µ the exciton mobility $\mu = D/k_B T$ with the Boltzmann constant $k_B$ and the temperature T. This equation is solved with the Pdepe solver of Matlab. In Figure 3d, in the case without any funnelling (D = 0) the exciton density remains a Gaussian distribution and is strongly perturbed at low r when the funnelling is applied ($D = 3.\,10^{-5}\,\text{m}^2.\,\text{s}^{-1}$).

To investigate this further, we used a detailed model that takes into account material doping and electrostatic effects. Our solution for the system relied on the coupled Van Roosbroeck differential equations set, as found in the Supplementary Information. The drift-diffusion equation used previously to explain the Seebeck effect in exciton scattering was an approximation of this model. [74] Thanks to this model, we can accurately discuss the significance of scattering caused by Auger recombination in contrast to excitonic radiative decay. When $R_A$ equals $0.1\,\text{cm}^2/\text{s}$, the Auger recombination term significantly alters the exciton concentration and funneling effect throughout the sample, ultimately reducing them. It appears crucial to take into account the Auger recombination of electrons when examining the diffusion of 2D excitons at ambient temperature. In our study, we focus on modelling photocreated species and solely account for neutral excitons. The presence of free charges is not accounted for, and we assume their uniform distribution in space. However, we observe that employing similar equations and parameters make it feasible to investigate the transport of free charges and their recombination under ambient conditions. It has previously been observed that the free charge funnelling is prevalent in the system at ambient conditions. It is important to note that the diffusion and significance of the funnel effect in this type of indentation experiment are dependent on multiple parameters, and specific development is necessary to address each effect individually.

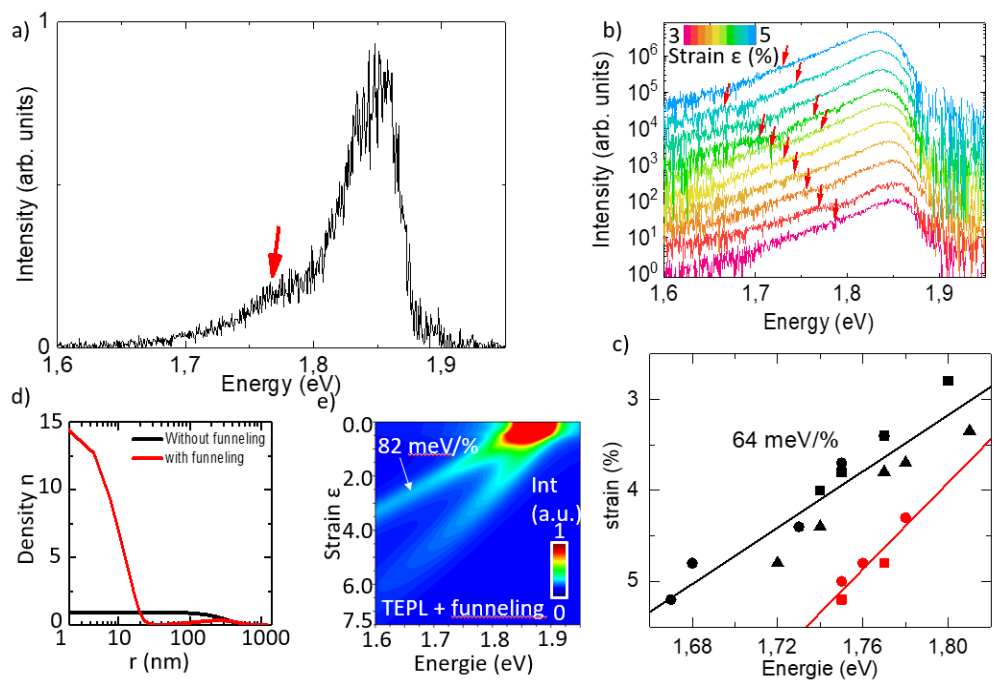

**Figure 5: TEPL signals show resurgences and PL signal under the tip.** a) and b) Examples of TEPL signals where a small satellite peak emerges at low energy, indicated by the red arrows. c) A compilation of all shoulder peaks at corresponding energy position and strain for different indentation measurements. Two groups of points are observed, which have been fitted with a linear curve. The black curve exhibits a slope of 64 meV/% elongation, which is comparable to the result obtained from simulations in e) d) The density of excitons along the sample with and without the funnelling effect and b) The PL response in the TEPL region, considering their dependence on energy and maximum strain (defined at the central position) with or without funnelling effect. A direct consequence of funnelling is to enhance and highlight a peak with a shift of 82 meV/% of maximum strain.

It should be noted that the experiment has verified the theoretical bandgap variation of 88meV per % elongation via elongation measurements of a 2D material subjected to a capacitive force. In addition, if the maximum elongation at the apex of the tip was eased by folds, wrinkles or other techniques, and decreased by a factor of two, for example, the shoulder TEPL signal, which moves quickly with elongation, could no longer be accommodated by a model. This reinforces the indentation model employed in this experiment. In other words, the measurement of the shoulder peak signal confirms the high strain of 10% applied to the ternary 2D and our model.

**Conclusion**

In the pursuit of comprehending novel phenomena and devising innovative apparatus pertaining to excitonic scattering, suitable methods are required for extracting the local optical response of a 2D material experiencing very high strain. Employing a TEPL approach, the optical response of single-layered $WS_{1.3}Se_{0.7}$ suspended membranes has been measured whilst undergoing local strain around 10%, which is one of the highest value reports so far in semiconducting 2D materials. This high strain is facilitated by the mechanical pressure imparted by the TEPL probe's tip. The nano-resolved photoluminescence signal at the tip end, which is also the location of highest deformation and strongest strain gradient, shifts significantly in energy. The presence of a shoulder peak has been measured which have a very high response, in peak position, to the membrane deflection. The experimental results are compared with models of 2D membrane deformation, simulation of the electronic band structure under strong deformation and exciton scattering/drifting in 2D materials, allowing a better understanding of the physics underlying the optical response in these devices. To our knowledge, optical measurements under tuneable and elevated strain in ambient conditions on 2D materials have not been reported before. The AFM-Raman/PL system is an ideal tool for studying vertical heterostructures to fabricate quantum source emitters. In the previous section, it was observed that tip-enhanced photoluminescence and the funnel have a positive impact on the enhancement of the optical signal of the tip-stressed 2D material in the experimental setup. This discussion suggests several diverse effects that must be considered in future experiments to accurately establish the contribution of the funnel effect and the various methods for enhancing this signal. Ternary 2D material are also interesting for in- and out-of-plane electrical polarization and exhibit Rashba effect and maybe strong piezoelectricity. [75] In the future, the control of mechanical deformation and emitted light of ternary 2D material is a way to study deeply the piezoelectric in ternary 2D materials.

**Supplementary material:** It includes additional TEPL measurements, tip width measurements and NanoPL distribution calibration. We also describe our model for the tip indentation and the DFT calculations for the band gap-strain relation.

**Acknowledgments:** We thank the fruitful discussion with Andrey Krayev. The work was supported by French grants ANR ANETHUM (ANR-19-CE24-0021), ANR Deus-nano (ANR-19-CE42-0005), ANR 2DHeco (ANR-20-CE05-0045) and by the French Renatech network.

**Data avaibility:** The data that support the findings of this study are available from the corresponding author upon reasonable request.

**Bibliography**

[1]   S. Deng, A. V. Sumant, V. Berry, *Nano Today* **2018**, *22*, 14.

[2]   Z. Dai, L. Liu, Z. Zhang, *Advanced Materials* **2019**, *31*, 1805417.

[3]   Strain-tuning of the electronic, optical, and vibrational properties of two-dimensional crystals: Applied Physics Reviews: Vol 8, No 2, https://aip-scitation-org.proxy.scd.u-psud.fr/doi/10.1063/5.0037852, accessed: Sep., 2021.

[4]   C. R. Zhu, G. Wang, B. L. Liu, X. Marie, X. F. Qiao, X. Zhang, X. X. Wu, H. Fan, P. H. Tan, T. Amand, B. Urbaszek, *Phys. Rev. B* **2013**, *88*, 121301.

[5]   H. Li, A. W. Contryman, X. Qian, S. M. Ardakani, Y. Gong, X. Wang, J. M. Weisse, C. H. Lee, J. Zhao, P. M. Ajayan, J. Li, H. C. Manoharan, X. Zheng, *Nat Commun* **2015**, *6*, 7381.

[6]  M. Zeng, J. Liu, L. Zhou, R. G. Mendes, Y. Dong, M.-Y. Zhang, Z.-H. Cui, Z. Cai, Z. Zhang, D. Zhu, T. Yang, X. Li, J. Wang, L. Zhao, G. Chen, H. Jiang, M. H. Rümmeli, H. Zhou, L. Fu, *Nat. Mater.* **2020**, *19*, 528.

[7]  J. Feng, X. Qian, C.-W. Huang, J. Li, *Nature Photon* **2012**, *6*, 866.

[8]  A. V. Tyurnina, D. A. Bandurin, E. Khestanova, V. G. Kravets, M. Koperski, F. Guinea, A. N. Grigorenko, A. K. Geim, I. V. Grigorieva, *ACS Photonics* **2019**, *6*, 516.

[9]  N. V. Proscia, Z. Shotan, H. Jayakumar, P. Reddy, C. Cohen, M. Dollar, A. Alkauskas, M. Doherty, C. A. Meriles, V. M. Menon, *Optica, OPTICA* **2018**, *5*, 1128.

[10] C. Palacios-Berraquero, D. M. Kara, A. R.-P. Montblanch, M. Barbone, P. Latawiec, D. Yoon, A. K. Ott, M. Loncar, A. C. Ferrari, M. Atatüre, *Nat Commun* **2017**, *8*, 15093.

[11] A. Branny, S. Kumar, R. Proux, B. D. Gerardot, *Nat Commun* **2017**, *8*, 15053.

[12] K. Parto, K. Banerjee, G. Moody, *arXiv:2009.07315 [physics, physics:quant-ph]* **2020**.

[13] K. F. Mak, K. He, C. Lee, G. H. Lee, J. Hone, T. F. Heinz, J. Shan, *Nature Mater* **2013**, *12*, 207.

[14] S. Bertolazzi, J. Brivio, A. Kis, *ACS Nano* **2011**, *5*, 9703.

[15] R. Zhang, V. Koutsos, R. Cheung, *Appl. Phys. Lett.* **2016**, *108*, 042104.

[16] M. G. Harats, J. N. Kirchhof, M. Qiao, K. Greben, K. I. Bolotin, *Nat. Photonics* **2020**, *14*, 324.

[17] D. Lloyd, X. Liu, J. W. Christopher, L. Cantley, A. Wadehra, B. L. Kim, B. B. Goldberg, A. K. Swan, J. S. Bunch, *Nano Lett.* **2016**, *16*, 5836.

[18] A. Castellanos-Gomez, R. Roldán, E. Cappelluti, M. Buscema, F. Guinea, H. S. J. van der Zant, G. A. Steele, *Nano Lett.* **2013**, *13*, 5361.

[19] J. Lee, S. J. Yun, C. Seo, K. Cho, T. S. Kim, G. H. An, K. Kang, H. S. Lee, J. Kim, *Nano Lett.* **2021**, *21*, 43.

[20] Strained Bubbles in van der Waals Heterostructures as Local Emitters of Photoluminescence with Adjustable Wavelength | ACS Photonics, https://pubs-acs-org.inp.bib.cnrs.fr/doi/10.1021/acsphotonics.8b01497, accessed: Sep., 2021.

[21] D. F. C. Leon, Z. Li, S. W. Jang, C.-H. Cheng, P. B. Deotare, *Applied Physics Letters* **2018**, *113*, 252101.

[22] H. Moon, G. Grosso, C. Chakraborty, C. Peng, T. Taniguchi, K. Watanabe, D. Englund, *Nano Lett.* **2020**, *20*, 6791.

[23] K. Datta, Z. Lyu, Z. Li, T. Taniguchi, K. Watanabe, P. B. Deotare, *Nat. Photon.* **2022**, *16*, 242.

[24] R. Rosati, S. Brem, R. Perea-Causín, R. Schmidt, I. Niehues, S. M. de Vasconcellos, R. Bratschitsch, E. Malic, *2D Mater.* **2020**, *8*, 015030.

[25] H. Lee, Y. Koo, J. Choi, S. Kumar, H.-T. Lee, G. Ji, S. H. Choi, M. Kang, K. K. Kim, H.-R. Park, H. Choo, K.-D. Park, *Science Advances* **2022**, *8*, eabm5236.

[26] R. Rosati, R. Schmidt, S. Brem, R. Perea-Causín, I. Niehues, J. Kern, J. A. Preuß, R. Schneider, S. Michaelis de Vasconcellos, R. Bratschitsch, E. Malic, *Nat Commun* **2021**, *12*, 7221.

[27] R. J. Gelly, D. Renaud, X. Liao, B. Pingault, S. Bogdanovic, G. Scuri, K. Watanabe, T. Taniguchi, B. Urbaszek, H. Park, M. Lončar, *Nat Commun* **2022**, *13*, 232.

[28] T. M. G. Mohiuddin, A. Lombardo, R. R. Nair, A. Bonetti, G. Savini, R. Jalil, N. Bonini, D. M. Basko, C. Galiotis, N. Marzari, K. S. Novoselov, A. K. Geim, A. C. Ferrari, *Phys. Rev. B* **2009**, *79*, 205433.

[29] S. Huang, G. Zhang, F. Fan, C. Song, F. Wang, Q. Xing, C. Wang, H. Wu, H. Yan, *Nat Commun* **2019**, *10*, 2447.

[30] J. Chaste, I. Hnid, L. Khalil, C. Si, A. Durnez, X. Lafosse, M.-Q. Zhao, A. T. C. Johnson, S. Zhang, J. Bang, A. Ouerghi, *ACS Nano* **2020**, *14*, 13611.

[31] K.-A. N. Duerloo, Y. Li, E. J. Reed, *Nat Commun* **2014**, *5*, 4214.

[32] Hierarchy of graphene wrinkles induced by thermal strain engineering: Applied Physics Letters: Vol 103, No 25, https://aip.scitation.org/doi/10.1063/1.4857115, accessed: Sep., 2021.

[33] Intrinsic Properties of Suspended MoS2 on SiO2/Si Pillar Arrays for Nanomechanics and Optics | ACS Nano, https://pubs.acs.org/doi/10.1021/acsnano.7b07689, accessed: May, 2021.

[34] B. Liu, Q. Liao, X. Zhang, J. Du, Y. Ou, J. Xiao, Z. Kang, Z. Zhang, Y. Zhang, *ACS Nano* **2019**, *13*, 9057.

[35] E. Blundo, C. Di Giorgio, G. Pettinari, T. Yildirim, M. Felici, Y. Lu, F. Bobba, A. Polimeni, *Advanced Materials Interfaces* **2020**, *7*, 2000621.

[36] A. B. G. Trabelsi, F. V. Kusmartsev, B. J. Robinson, A. Ouerghi, O. E. Kusmartseva, O. V. Kolosov, R. Mazzocco, M. B. Gaifullin, M. Oueslati, *Nanotechnology* **2014**, *25*, 165704.

[37] Y. Xie, J. Lee, H. Jia, P. X.-L. Feng, in *2019 20th International Conference on Solid-State Sensors, Actuators and Microsystems Eurosensors XXXIII (TRANSDUCERS EUROSENSORS XXXIII)*, **2019**, pp. 254–257.

[38] M. Goldsche, G. J. Verbiest, T. Khodkov, J. Sonntag, N. von den Driesch, D. Buca, C. Stampfer, *Nanotechnology* **2018**, *29*, 375301.

[39] F. Guan, P. Kumaravadivel, D. V. Averin, X. Du, *Applied Physics Letters* **2015**, *107*, 193102.

[40] S. Lou, Y. Liu, F. Yang, S. Lin, R. Zhang, Y. Deng, M. Wang, K. B. Tom, F. Zhou, H. Ding, K. C. Bustillo, X. Wang, S. Yan, M. Scott, A. Minor, J. Yao, *Nano Lett.* **2018**, *18*, 1819.

[41] Coherent, atomically thin transition-metal dichalcogenide superlattices with engineered strain, https://www.science.org/doi/10.1126/science.aao5360?url_ver=Z39.88-2003&rfr_id=ori:rid:crossref.org&rfr_dat=cr_pub%20%200pubmed, accessed: Sep., 2021.

[42] Z. Ben Aziza, H. Henck, D. Di Felice, D. Pierucci, J. Chaste, C. H. Naylor, A. Balan, Y. J. Dappe, A. T. C. Johnson, A. Ouerghi, *Carbon* **2016**, *110*, 396.

[43] J. Chaste, A. Missaoui, A. Saadani, D. Garcia-Sanchez, D. Pierucci, Z. Ben Aziza, A. Ouerghi, *ACS Appl. Nano Mater.* **2018**, *1*, 6752.

[44] J. N. Kirchhof, K. Weinel, S. Heeg, V. Deinhart, S. Kovalchuk, K. Höflich, K. I. Bolotin, *Nano Lett.* **2021**, *21*, 2174.

[45] D. Davidovikj, M. Poot, S. J. Cartamil-Bueno, H. S. J. van der Zant, P. G. Steeneken, *Nano Lett.* **2018**, *18*, 2852.

[46] S. Manzeli, A. Allain, A. Ghadimi, A. Kis, *Nano Lett.* **2015**, *15*, 5330.

[47] T. P. Darlington, C. Carmesin, M. Florian, E. Yanev, O. Ajayi, J. Ardelean, D. A. Rhodes, A. Ghiotto, A. Krayev, K. Watanabe, T. Taniguchi, J. W. Kysar, A. N. Pasupathy, J. C. Hone, F. Jahnke, N. J. Borys, P. J. Schuck, *Nat. Nanotechnol.* **2020**, *15*, 854.

[48] Y. Koo, Y. Kim, S. H. Choi, H. Lee, J. Choi, D. Y. Lee, M. Kang, H. S. Lee, K. K. Kim, G. Lee, K.-D. Park, *Advanced Materials* **2021**, *33*, 2008234.

[49] J.-T. Huang, B. Bai, Y.-X. Han, P.-Y. Feng, X.-J. Wang, X.-Z. Li, G.-Y. Huang, H.-B. Sun, *ACS Nano* **2023**, DOI: 10.1021/acsnano.3c06102.

[50] M. Gastaldo, J. Varillas, Á. Rodríguez, M. Velický, O. Frank, M. Kalbáč, *npj 2D Mater Appl* **2023**, *7*, 1.

[51] A. Chiout, F. Correia, M.-Q. Zhao, A. T. C. Johnson, D. Pierucci, F. Oehler, A. Ouerghi, J. Chaste, *Appl. Phys. Lett.* **2021**, *119*, 173102.

[52] X. Duan, C. Wang, Z. Fan, G. Hao, L. Kou, U. Halim, H. Li, X. Wu, Y. Wang, J. Jiang, A. Pan, Y. Huang, R. Yu, X. Duan, *Nano Lett.* **2016**, *16*, 264.

[53] C. Ernandes, L. Khalil, H. Almabrouk, D. Pierucci, B. Zheng, J. Avila, P. Dudin, J. Chaste, F. Oehler, M. Pala, F. Bisti, T. Brulé, E. Lhuillier, A. Pan, A. Ouerghi, *npj 2D Materials and Applications* **2021**, *5*, 1.

[54] P. Wang, S. Song, A. Najafi, C. Huai, P. Zhang, Y. Hou, S. Huang, H. Zeng, *ACS Nano* **2020**, DOI: 10.1021/acsnano.0c02838.

[55] K. Kim, M. Yankowitz, B. Fallahazad, S. Kang, H. C. P. Movva, S. Huang, S. Larentis, C. M. Corbet, T. Taniguchi, K. Watanabe, S. K. Banerjee, B. J. LeRoy, E. Tutuc, *Nano Lett.* **2016**, *16*, 1989.

[56] X. Ma, Q. Liu, D. Xu, Y. Zhu, S. Kim, Y. Cui, L. Zhong, M. Liu, *Nano Lett.* **2017**, *17*, 6961.

[57] Z. He, Z. Han, J. Yuan, A. M. Sinyukov, H. Eleuch, C. Niu, Z. Zhang, J. Lou, J. Hu, D. V. Voronine, M. O. Scully, *Science Advances* **2019**, *5*, eaau8763.

[58] Y. Okuno, O. Lancry, A. Tempez, C. Cairone, M. Bosi, F. Fabbri, M. Chaigneau, *Nanoscale* **2018**, *10*, 14055.

[59] C. Tang, Z. He, W. Chen, S. Jia, J. Lou, D. V. Voronine, *Phys. Rev. B* **2018**, *98*, 041402.

[60] Z. Zhang, A. C. De Palma, C. J. Brennan, G. Cossio, R. Ghosh, S. K. Banerjee, E. T. Yu, *Phys. Rev. B* **2018**, *97*, 085305.

[61] P. K. Sahoo, H. Zong, J. Liu, W. Xue, X. Lai, H. R. Gutiérrez, D. V. Voronine, *Opt. Mater. Express, OME* **2019**, *9*, 1620.

[62] C. Jin, A. Davoodabadi, J. Li, Y. Wang, T. Singler, *Journal of the Mechanics and Physics of Solids* **2017**, *100*, 85.

[63] N. M. Bhatia, W. Nachbar, *AIAA Journal* **1968**, *6*, 1050.

[64] D. Vella, B. Davidovitch, *Soft Matter* **2017**, *13*, 2264.

[65] M. R. Begley, T. J. Mackin, *Journal of the Mechanics and Physics of Solids* **2004**, *52*, 2005.

[66] E. Schwerin, *ZAMM - Journal of Applied Mathematics and Mechanics / Zeitschrift für Angewandte Mathematik und Mechanik* **1929**, *9*, 482.

[67] M. Annamalai, S. Mathew, M. Jamali, D. Zhan, M. Palaniapan, *J. Micromech. Microeng.* **2012**, *22*, 105024.

[68] C. Lee, X. Wei, J. W. Kysar, J. Hone, *Science* **2008**, *321*, 385.

[69] F. Zeng, W.-B. Zhang, B.-Y. Tang, *Chinese Phys. B* **2015**, *24*, 097103.

[70] J. Kang, S. Tongay, J. Zhou, J. Li, J. Wu, *Appl. Phys. Lett.* **2013**, *102*, 012111.

[71] A. Falin, M. Holwill, H. Lv, W. Gan, J. Cheng, R. Zhang, D. Qian, M. R. Barnett, E. J. G. Santos, K. S. Novoselov, T. Tao, X. Wu, L. H. Li, *ACS Nano* **2021**, *15*, 2600.

[72] D. Thureja, A. Imamoglu, T. Smoleński, I. Amelio, A. Popert, T. Chervy, X. Lu, S. Liu, K. Barmak, K. Watanabe, T. Taniguchi, D. J. Norris, M. Kroner, P. A. Murthy, *Nature* **2022**, *606*, 298.

[73] P. Lin, L. Zhu, D. Li, L. Xu, C. Pan, Z. Wang, *Advanced Functional Materials* **2018**, *28*, 1802849.

[74] S. Park, Polarisation Resolved Photoluminescence in van Der Waals Heterostructures, These de doctorat, Institut polytechnique de Paris, **2022**.

[75] J. Zribi, D. Pierucci, F. Bisti, B. Zheng, J. Avila, L. Khalil, C. Ernandes, J. Chaste, F. Oehler, M. Pala, T. Maroutian, I. Hermes, E. Lhuillier, A. Pan, A. Ouerghi, *Nanotechnology* **2022**, *34*, 075705.

# Supplementary information

# High Strain Engineering of a Suspended WSSe Monolayer Membrane by Indentation and Measured by Tip-enhanced Photoluminescence.

Anis Chiout [1], Agnès Tempez [2], Thomas Carlier [2], Marc Chaigneau [2], Fabian Cadiz[3], Alistair Rowe[3], Biyuan Zheng [4], Anlian Pan [4], Marco Pala [1], Fabrice Oehler [1], Abdelkarim Ouerghi [1], Julien Chaste [1*]

[1] Université Paris-Saclay, CNRS, Centre de Nanosciences et de Nanotechnologies, 91120, Palaiseau, France.
[2] HORIBA France SAS, Passage Jobin Yvon, 91120 Palaiseau, France
[3] Key Laboratory for Micro-Nano Physics and Technology of Hunan Province, State Key Laboratory of Chemo/Biosensing and Chemometrics, and College of Materials Science and Engineering, Hunan University, 410082 Changsha, Hunan, China.
[3] Laboratoire de Physique de la Matière Condensée, CNRS, Ecole Polytechnique, Institut Polytechnique de Paris, 91120 Palaiseau, France.

* Corresponding author: julien.chaste@universite-paris-saclay.fr

## Table des matières

# S1) Suspended membranes of monolayer WSSe made by CVD.

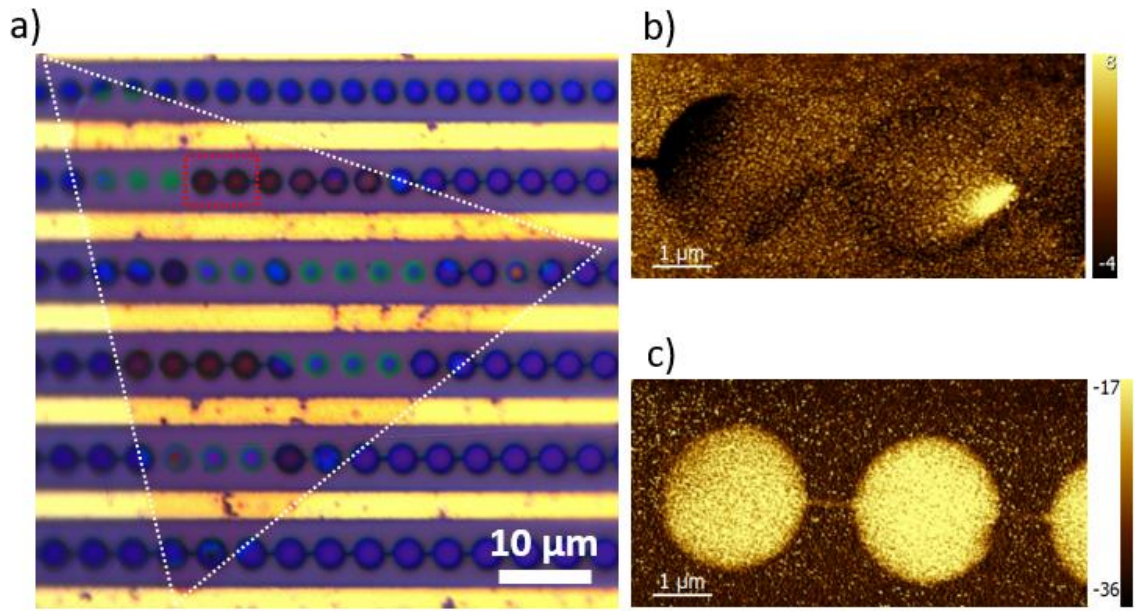

**Figure S1a: Optical image and topography of the WSSe flake.** a) A large triangular flake of WSSe was deposited on a pre-patterned structure with an interdigitated gold contact and a multiple array of designed holes in the $SiO_2$. All holes are connected by a small 200nm wide trench to evacuate air in the cavity during indentation. Atomic force microscopy; topography (b) and phase (c) of the suspended membrane tested for measurements before the indentation measurement. The membrane is completely flat with no ripples or defects.

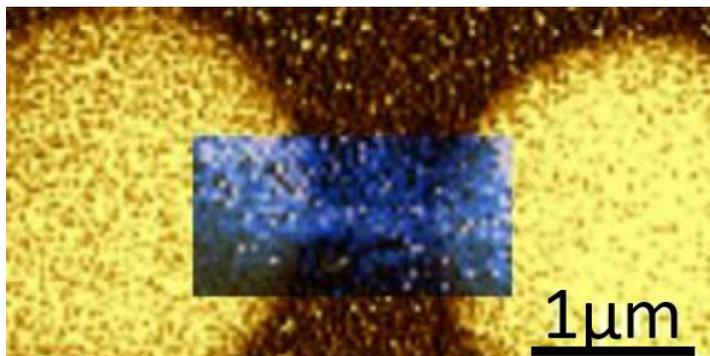

**Figure S1b: AFM signal and tip-enhanced signal of the WSSe flake.** Superposition of the atomic force microscopy image (phase) and the tip-enhanced photoluminescence (TEPL) amplitude (blue image in the center). The TEPL signal is higher in the suspended parts of the WSSe membrane. We obtain a lateral resolution well below hundreds of nm. It is possible to observe the narrow trench between the two drums.

# S2) Measure of the tip radius and TEPL distribution width.

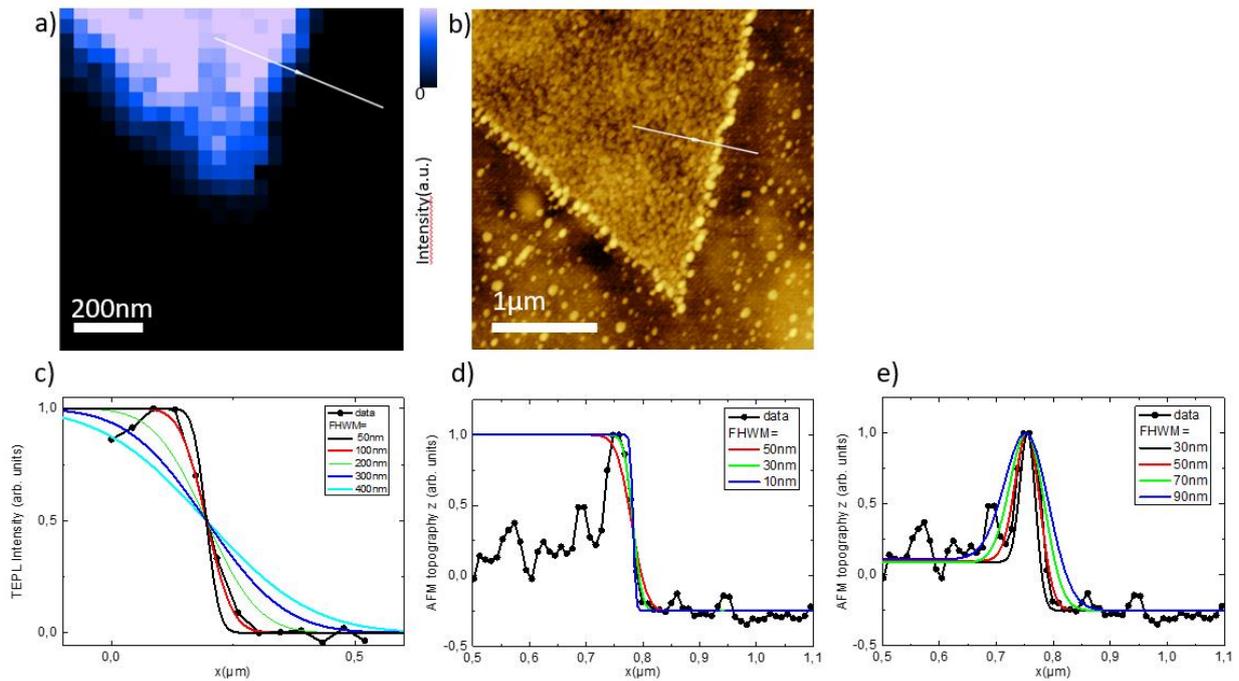

**Figure S2: Photoluminescence image enhanced by the tip of a WSSe flake on SiO$_2$.** A large triangular flake of WSSe was deposited on top of SiO$_2$. In (a), we measured the TEPL signal at the corner of a 2D flake shown in the AFM image (b). The TEPL signal is obtained here by measuring the difference in photoluminescence with and without the metal tip. In (c), we compared the TEPL signal along the edge (white line in (a)) with our model. The approximate model is a convoluted Heaviside step with a 2D Gaussian distribution. The full width at half maximum (FWHM) of the TEPL signal is 90nm. (d) The same treatment was performed for the AFM topography to extract the tip radius, we obtain 30nm for the FWHM. In (e), we consider a more accurate rectangular function of width=10nm in addition to the Heaviside step to improve the fit. The width is 50nm. As a result, we determine that the AFM tip used for indentation and TEPL measurements has a radius at the apex of 40nm.

## S3) Indentation and deflection measurement.

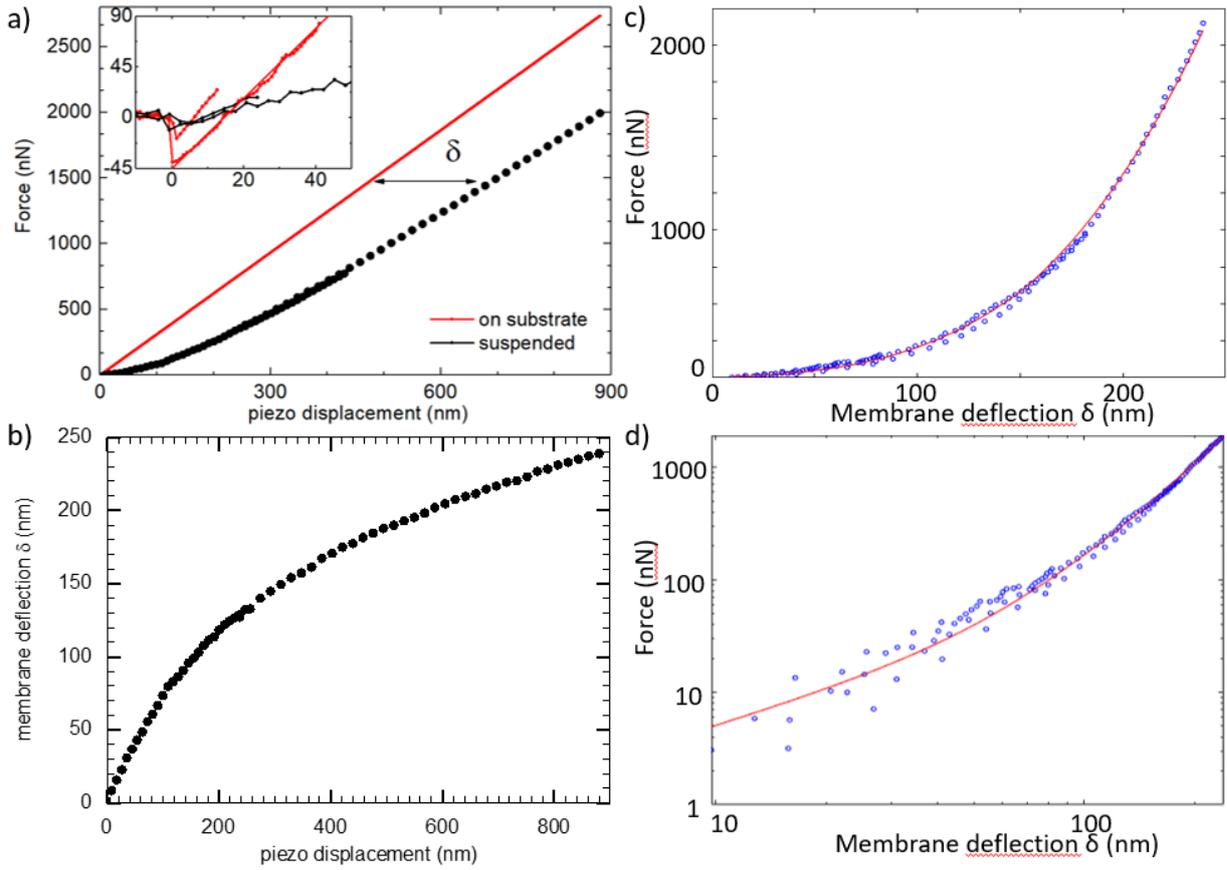

**Figure S3a: Calibration of the indentation for the AFM tip (tip spring constant $k_{tip}$=3.11N/m)** a) The force, F, applied to the 2D membrane as function of the piezo displacement($z_{piezo}$) measured on the 2D material with substrate (red) and on the suspended part (red). In inset, a zoom for the small deflection values. For the suspended part, we recover a spring constant of $k_{tip}$=3.11N/m with F=$k_{tip}$.$z_{piezo}$ which was calibrated independently. b) membrane deflection δ in function of the piezo displacement. The membrane deflection is measured by the shift in piezo displacement at each force between red and black curve in (a). In c) and d), the force in function of the membrane deflection δ in linear and log scale. With the theoretical fit curve in red.

We use an empirical model to describe the relation between the piezostage movement $z_{piezo}$ and the membrane deflection (Figure S3a,b).

$$\delta = 422.78 - 148.36 \times e^{\frac{-(z_{piezo}+47.58)}{151.79}} - 321.60 \times e^{\frac{-(z_{piezo}+47.58)}{1649.09}}$$

**Tip model**

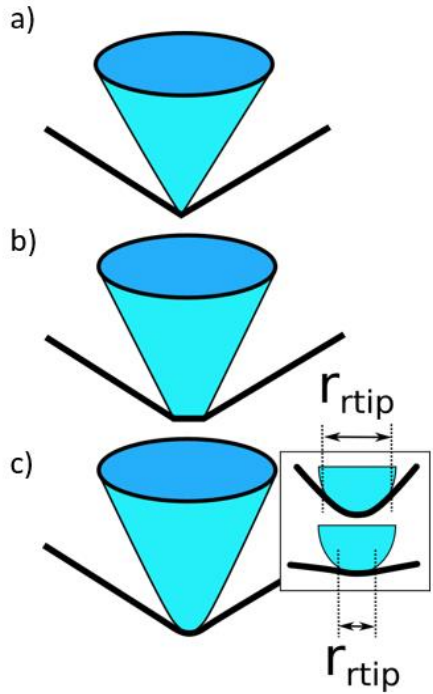

**Figure S3b: The different tip shape.** a) To a first approximation, the radius of the tip can be considered very small compared to the radius of the membrane. ($r_{tip} \sim 0 \ll R$), it corresponds to the standard model of Schwerin type. b) the tip has a finite flat apex of radius $r_{tip} = 20nm$. The contact surface is fixed. c) The tip as a round surface with some radius curvature. In this model, the radius of the contact surface, $r_{contact}$, is dependant of the strain.

In order to obtain a contact surface $r_{contact} = r_{tip} = 20nm$ for very small indentation ($\varepsilon < 0.5\%$), as measured experimentally, we have to fixed the radius curvature at 60nm. At maximum strain, the contact surface increase to $r_{contact} \sim 40nm$

**Model indentation Schwerin-type standard**[1] [2–5]

This model defined the indentation of a membrane with a small tip of zero radius

$$F = \sigma_{2D}.\pi\delta + E_{2D}.\frac{q_1^3}{R^2}.\delta^3$$

$\sigma_{2D}$ is the pretension term, R is the membrane radius. The stretching modulus is $E_{2D} = E.t$ with E the Young modulus and t=0.7nm the sample thickness, $q_1 = 1/(1.05 - 0.15.\nu - 0.16\nu^2)$. $\nu = 0.19$ is the Poisson ratio of the material.

F=indenter load
t=membrane thickness
E=Young modulus
$E_{2D}$=stretching modulus
$\varepsilon_0$=bi axial prestrain
$r_{tip}$= radius indenter

**Model indentation from Föppl–von-Kármán equations** [6]

This model describe the indentation of a membrane with a tip which is round at the end and has a small radius of curvature [6] ( limits for $0 \ll r \ll R$, large indentation ($\tilde{\delta} \gg 1$), $0 < \nu < 0.5$)
$q_2 = 0.867 + 0.2773.\nu + 0.8052.\nu^2$
We use dimensionless parameters $\tilde{\delta} = N_\delta.\delta$ and $\tilde{F} = N_F.F$ with $N_\delta = E_{2D}^{0.5}/(R.\sigma_{2D}^{0.5})$ and $N_F = E_{2D}^{0.5}/(R.\sigma_{2D}^{1.5})$

$$\frac{\tilde{F}}{\tilde{\delta}^3} = q_2 + \left(\frac{6}{2\pi(1+\nu)}\right)^{\frac{1}{3}} . q_2^{\frac{4}{3}} . \left(\frac{r_{tip}}{R}\right)^{\frac{2}{3}}$$

# S4) Strain at the center and along the membrane.

### Model strain at the center for Schwerin type model
The Schwerin type model is defined with a point load probe

The maximum stress and strain are:

$$\sigma_{max} = \sqrt{\frac{F.Et}{4\pi.r_{tip}}}$$

$$\varepsilon_{max} = \frac{\sigma_{max}}{Et} = \sqrt{\frac{F}{4\pi.r_{tip}.Et}}$$

### Model strain in $1/r$
If we consider the tip to be flat, model b in Figure S3b, simple approximation have consider a model with a $1/r$ dependence [7,8]

$$\begin{cases} \varepsilon(r) = \varepsilon_{max} \text{ when } r \leq r_{tip} \\ \varepsilon(r) = \frac{\varepsilon_{max}.r_{tip}}{r} \text{ when } r > r_{tip} \end{cases}$$

### Model strain along the membrane above 40nm from the tip center [9]
This model describes the strain along the membrane and far away from the tip. (small strain, $u(r > r_{tip}) \sim 0$).
In this model, the tip is rounded at the apex with a radius curvature. (model c in figure S3b)
u the radial displacement and w the transversal displacement $w(r = 0) \sim \delta$
Radial displacement is dominated by prestress $u = \varepsilon_0 r$

$$\varepsilon = \frac{du}{dr} + \frac{1}{2}.\left(\frac{dw}{dr}\right)^2 \approx \frac{1}{2}.\left(\frac{dw}{dr}\right)^2$$

If we define the parameter AA as;

$$AA = \frac{2}{3\pi} \cdot \frac{\left(\sqrt{9\pi^4 \cdot \left(\frac{F}{Etr_{tip2}}\right)^2 + 64\pi^6 \cdot \varepsilon_0^3} + 3\pi^2 \cdot \frac{F}{Etr_{tip2}}\right)^{\frac{2}{3}} - 4\pi^2 \cdot \varepsilon_0}{\left(\sqrt{9\pi^4 \cdot \left(\frac{F}{Etr_{tip2}}\right)^2 + 64\pi^6 \cdot \varepsilon_0^3} + 3\pi^2 \cdot \frac{F}{Etr_{tip2}}\right)^{\frac{1}{3}}}$$

For zero pre-strain, AA simplify to

$$AA = \left(\frac{16}{9\pi} \cdot \frac{F}{Etr_{tip2}}\right)^{\frac{1}{3}}$$

We are able to describe $w$ as;

$$w = AA \cdot r_{tip} \cdot \left(\left(\frac{R}{r_{tip2}}\right)^{\frac{3}{4}} - \left(\frac{r}{r_{tip2}}\right)^{\frac{3}{4}}\right)$$

From w, we deduce the strain $\varepsilon$ along the membrane. A good approximation of the function is a function $\varepsilon(r) \propto 1/\sqrt{r}$. This description seems accurate and describe well the strain in the membrane for distance above 40nm from the center.

On contrary, at the center it diverges from realistic values.
For small strain, in the ref [9], the authors explicitly gives a formula for the strain at the center;

$$\varepsilon(r = 0) \approx \frac{1}{6}\sqrt{\left(\frac{3}{2\pi}\right)\left(\frac{F}{Etr_{tip2}} + 6\pi\varepsilon_0^2\right)} - \frac{1}{2}\varepsilon_0$$

For a negligible $\varepsilon_0$, we obtain;

$$\varepsilon(r = 0) \approx \frac{1}{\sqrt{6}}\sqrt{\frac{F}{4\pi Etr_{tip2}}}$$

This value has a factor $\sqrt{6} \sim 2.5$ with the maximum strain obtain in the case of the Schwerin type model and of the strain model describe below for small r.

## Model strain along the membrane in the contact region

Model [10] and [11] are used in the contact region, where the membrane is in contact to the tip at any position $r < r_{contact}$

$$r_{contact} = r_{tip} \cdot \sqrt{\frac{3}{2\pi} \cdot \frac{F}{Etr_{tip2}} + 6\pi\varepsilon_0^2 - 3\varepsilon_0}$$

The stress at the contact is

$$N_c = \frac{F \cdot r_{tip2}}{2\pi \cdot r_{contact}^2}$$

The deflection u, w and strain $\varepsilon$ along the membrane are defined by;

$$u = r \cdot \left((1-\nu)\frac{N_c}{Et} + \frac{(1-\nu)}{16} \cdot \frac{r_{contact}^2}{r_{tip2}^2} - \frac{(3-\nu)}{16} \cdot \frac{r^2}{r_{tip2}^2}\right)$$

$$w = w_0 + \frac{r_{tip2}}{2} \cdot \left(\frac{r_{contact}^2}{r_{tip2}^2} - \frac{r^2}{r_{tip2}^2}\right)$$

$$\varepsilon = \frac{du}{dr} + \frac{1}{2} \cdot \left(\frac{dw}{dr}\right)^2$$

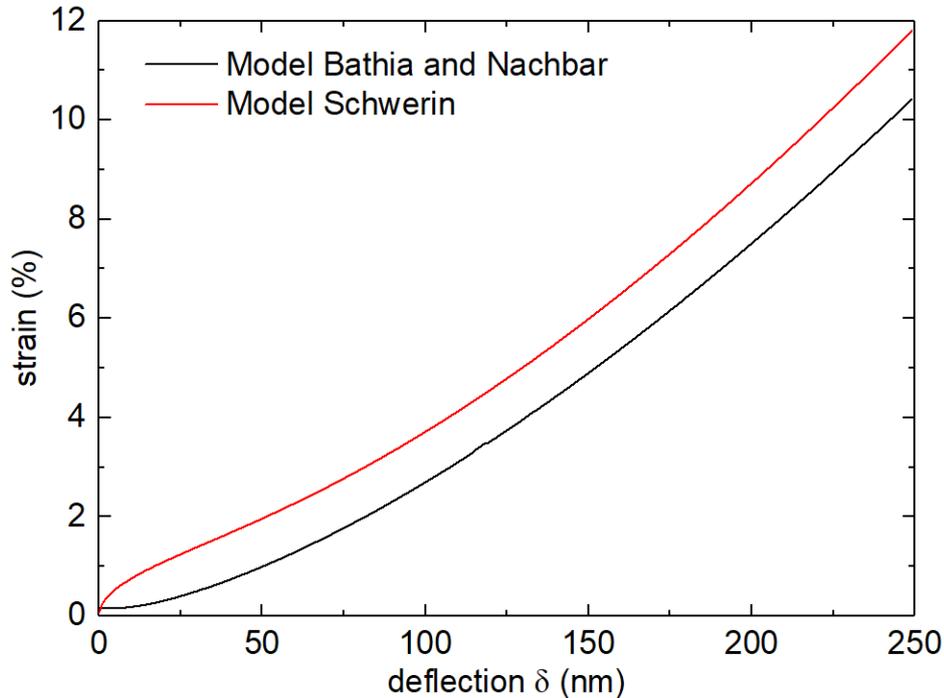

**Figure S4a: Strain at the tip center for different model in function of the indentation.** We plot the two model of indentation: the Bathia and Nachbar model (finite-sized and rounded probe) [10,11] and the standard model of Schwerin type. [1] [2–5] (point probe). We deduce from this a maximum strain 10.3% at the center of our membrane in our experiment.

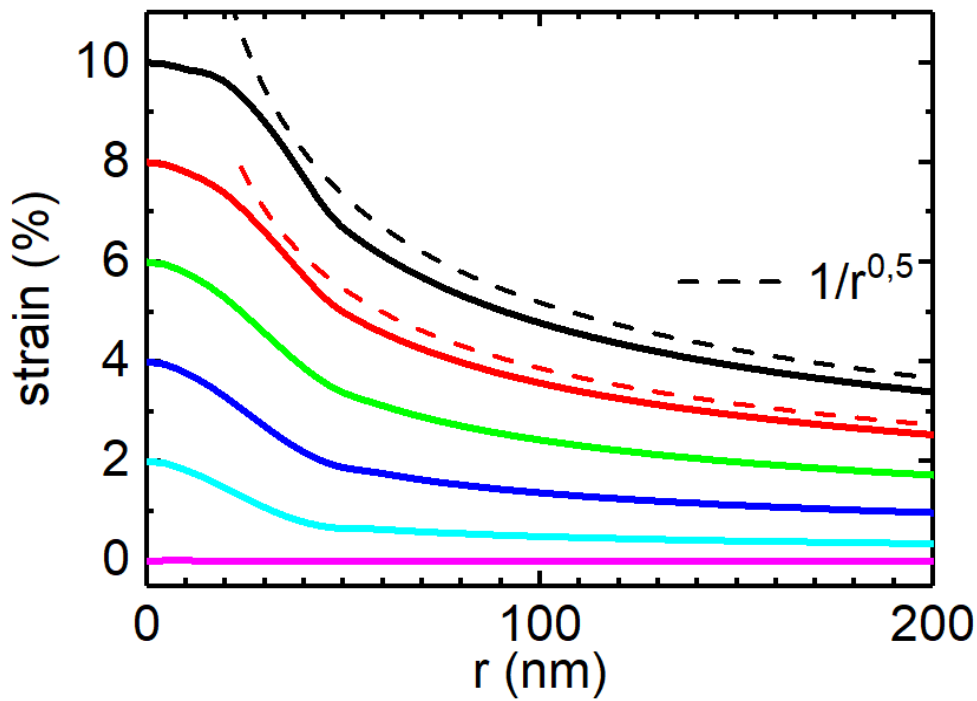

**Figure S4b: Strain along the membrane.** Model of strain along the radial direction of the circular membrane for different indentations. The curve has two limits, one closed to the contact region (r<20nm) and the other one for the suspended part (r>40nm). The dependence of the strain in this suspended part is closed to a $\varepsilon \propto 1/\sqrt{r}$ function.

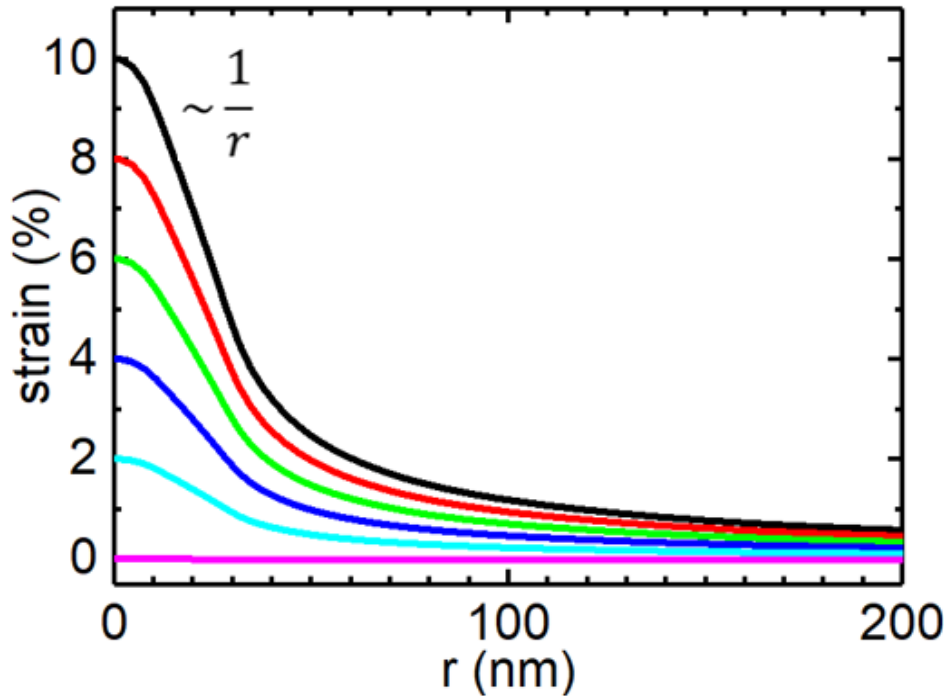

**Figure S4c: Strain along the membrane.** Model of strain along the radial direction of the circular membrane for different indentations. The model is a $\varepsilon \propto 1/r$ function which was smoothed in order to avoid some divergence in the calculation of the funneling effect.

## S5) Tip enhanced photoluminescence measurements

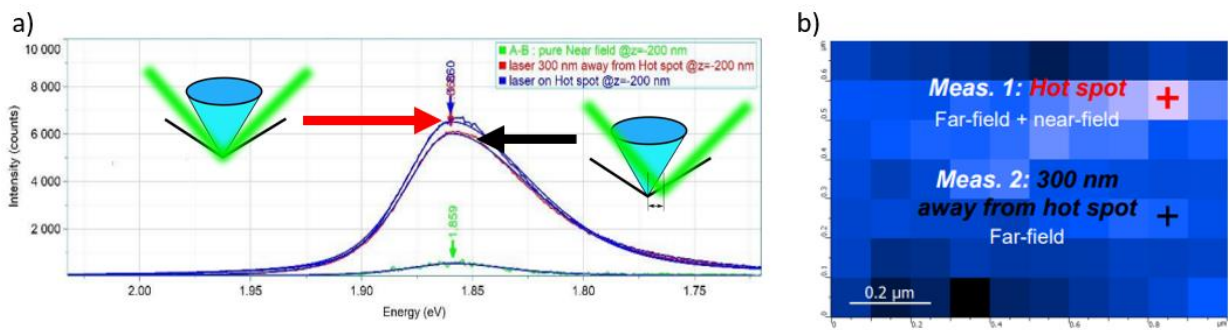

**Figure S5a: Tip-enhanced photoluminescence signal as a function of laser and tip spatial alignment.** a) PL signal at the hot spot (blue curve), when the laser and AFM tip are aligned to enhance the TEPL signal and PL signal when the laser and tip are misaligned by 300nm (red curve). In green, the signal obtained by simple subtraction of the two previous signals which contains mainly the TEPL signal (14% of the mains signal amplitude) b) Mapping of the TEPL signal as a function of the laser position relative to the tip position in the x and y directions. We can see the appearance of a hot spot around the tip.

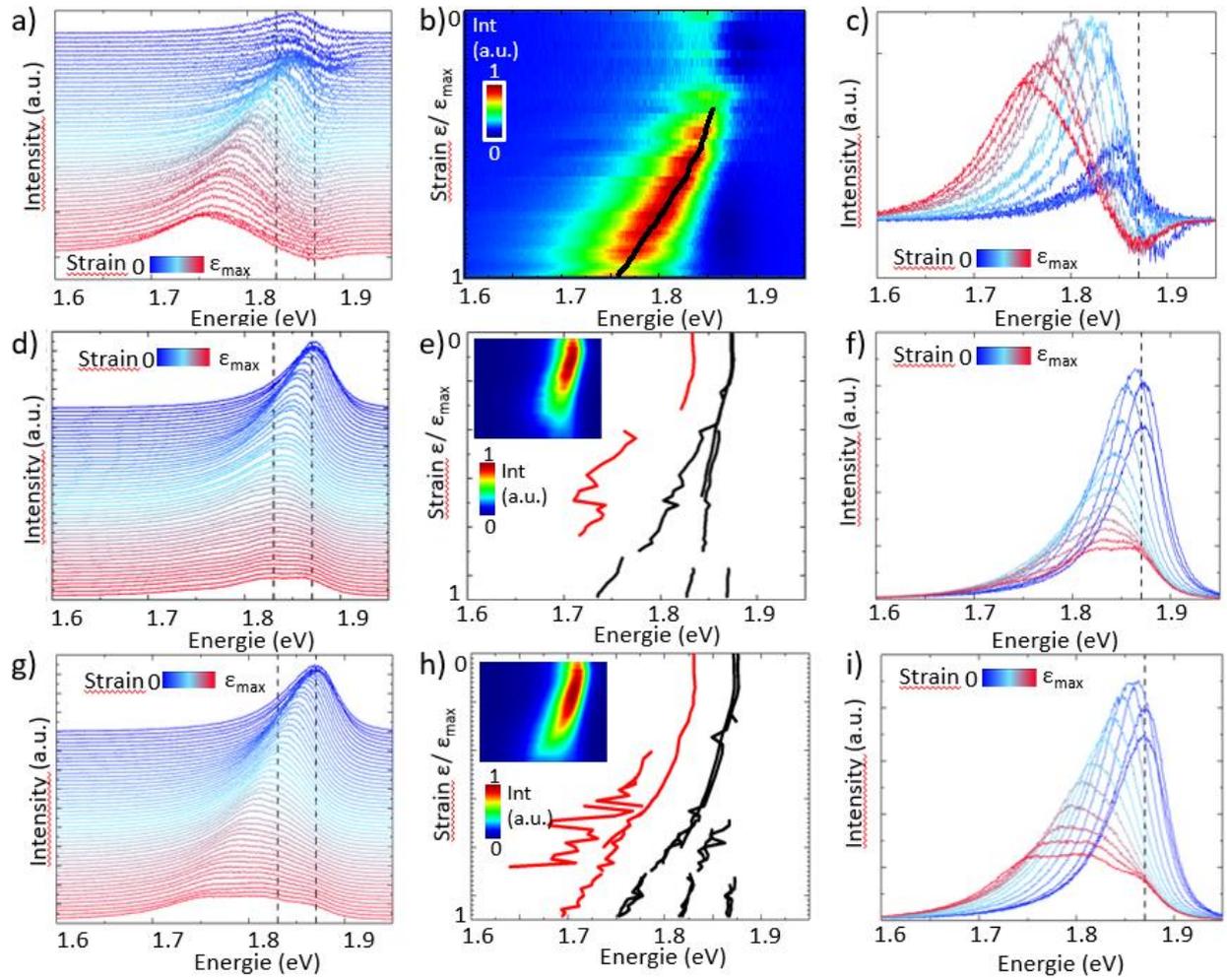

**Figure S5b: Optical image and topography of the WSSe flake.** A large triangular flake of WSSe has been deposited.

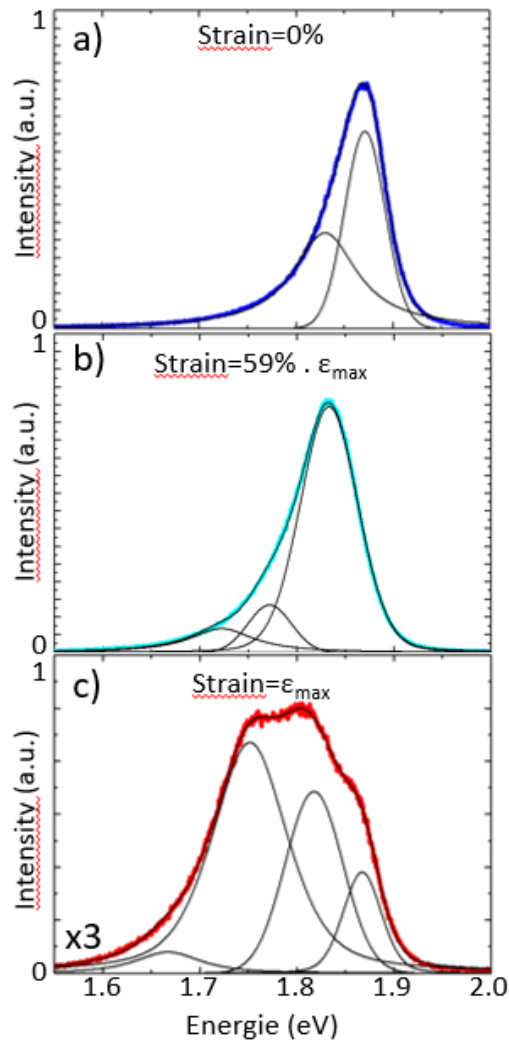

**Figure S5c: Photoluminescence of WSSe flake with and without stress.** a) PL measurement of a suspended membrane without applied stress. The fit is made with two peaks; a Lorentzian peak and a Gaussian peak. These peaks correspond to the A exciton and A' trion of the WSSe. b) and c) PL measurement for different peaks, it seems reasonable to fit the curves with an increasing number of peaks as the strain increases. Depending on the strain, we fit with two, three or four peaks.

The two main peak without strain are a Lorentzian peak center around 1.833eV (width=81.5meV, Amplitude=2289 counts) and a Gaussian peak at 1.873eV (width=47.7meV, Amplitude=5252 counts).

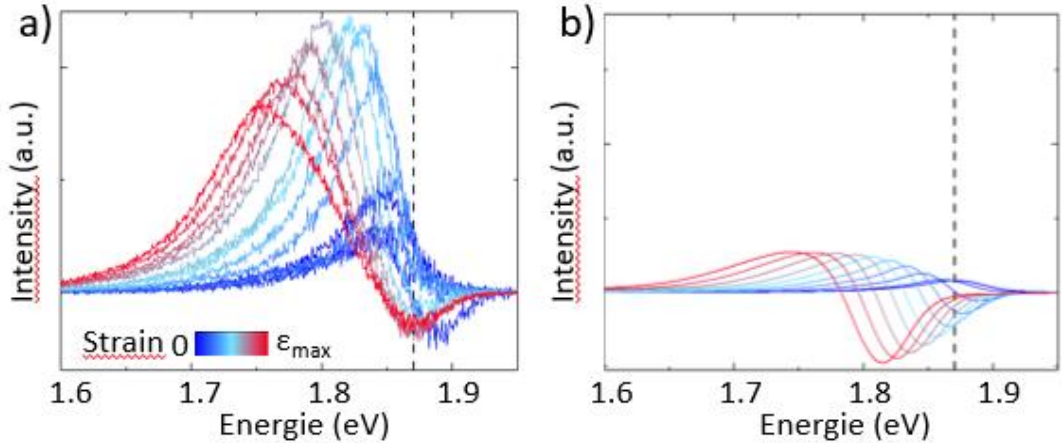

**Figure S5d: Photoluminescence of a WSSe flake - residual. of the far field signal.** Usually TEPL measurements are performed by subtracting the signal in presence and without the AFM tip, the far field signal being strictly the same between the two positions. Here, by experimental necessity, we measure the difference of the signal at two positions of the laser with respect to the tip. The distance of 300nm separating these two positions strongly reduces the TEPL signal when the alignment with the tip is not perfect. But we have to consider the small far-field signal difference between these two positions. In a) the experimentally measured difference and in b) the simulated difference of the far field signal between the two positions. We observe that a part of the signal is related to the far field residual and our measurement technique and can be extracted from our data. On the other hand, we can see the main contribution of the TEPL signal which appears in the measurement by subtracting the calculated signal from the experimental signal.

## S6) DFT calculation of WSSe, WSe$_2$ and WS$_2$ band structure vs strain

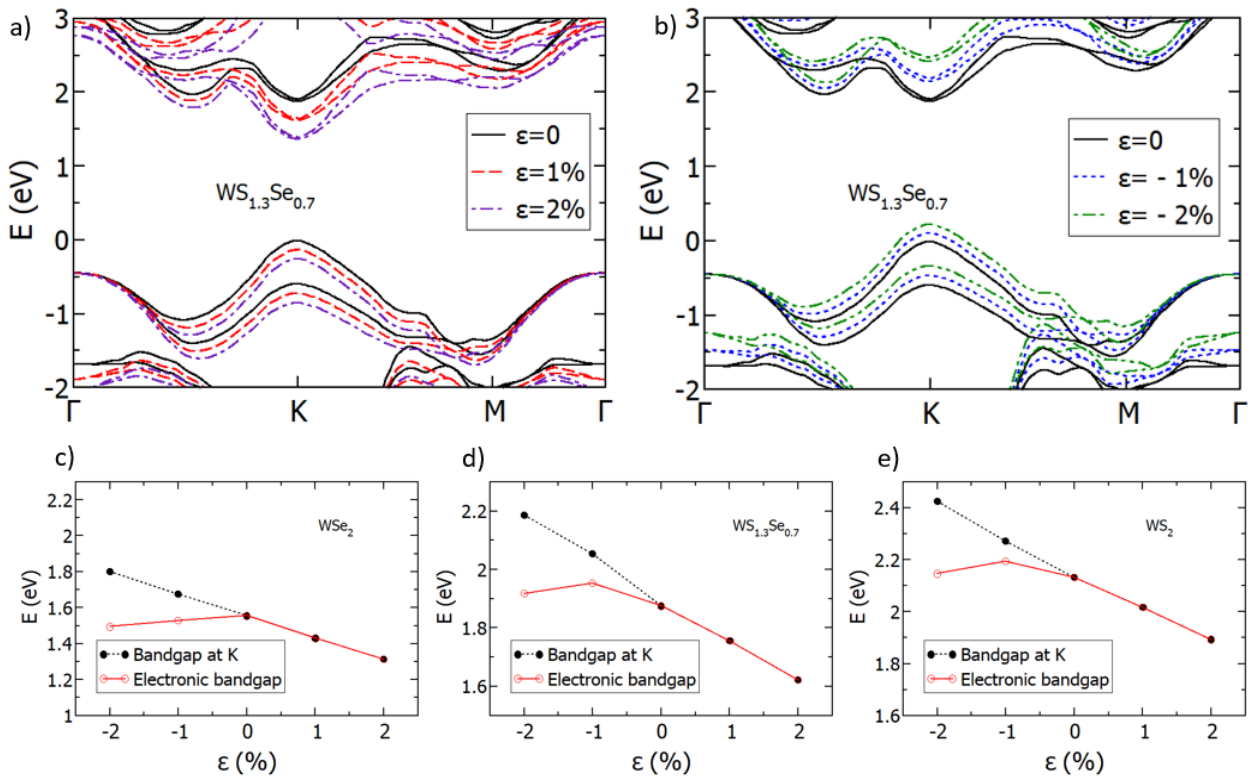

**Figure S6: Band structure evolution of different 2D materials monolayer under biaxial strain.** The band structure of WSSe have been obtained with density Functional theory (DFT) for different strain of elongation in a) and compression in b). The applied strain is an in-plane strain and applied in both direction, x and y. Evolution of the band gap in function of the strain c) for $WSe_2$, d) for WSSe, e) $WS_2$. Under compression, the material transit from a direct band gap semiconductor to an indirect band gap semiconductor for strain around -1%.

The slope of the band gap in function of the positive strain for WSSe is 125meV/% of elongation. The strain $\varepsilon$ is applied in both direction x,y as $\varepsilon = \varepsilon_x = \varepsilon_y$. We consider the radial strain to be $\varepsilon_r = \sqrt{\varepsilon_x^2 + \varepsilon_y^2} = \sqrt{2}\varepsilon$. It means the dependence of the band gap with the radial strain is 88meV per % of elongation.
Our model does not consider the excitonic energy, i.e. the additional energy necessary to link an electron-hole pair.

## S7) Optomechanical measurement and capacitive pull-in

It is possible to apply a uniform non-local strain on a suspended membrane with an electrostatic force. It implies to contact our 2D material with source-drain electrodes and a back-gate voltage. We measure the vibration and the strain in this suspended 2D materials with optomechanical measurements during a gate voltage sweep. We proceed to such measurements at low pressure. The photoluminescence response of the same membrane has also been measured at low pressure in function of the gate voltage. We observe a small shift of the peak position in function of the gate voltage which is attributed to the capacitive force applied to the membrane.
It is possible to confirm in Figure S7a that the electrostatic force $F=C'V_G^2$ and the strain related to this effect is symmetric with $V_G$, with $V_G$ the gate voltage and C' the derivative of the capacitance with the distance gate, membrane C'=dC/dz. In order to only consider and extract the effect of the capacitive pull in on the membrane from the photoluminescence measurements, we extract the symmetrical PL response with respect to $V_G$ from our data. The asymmetrical component is non null and is certainly related to a doping effect due to the gate voltage variation. The shifts are the same for the two components and around 5meV.

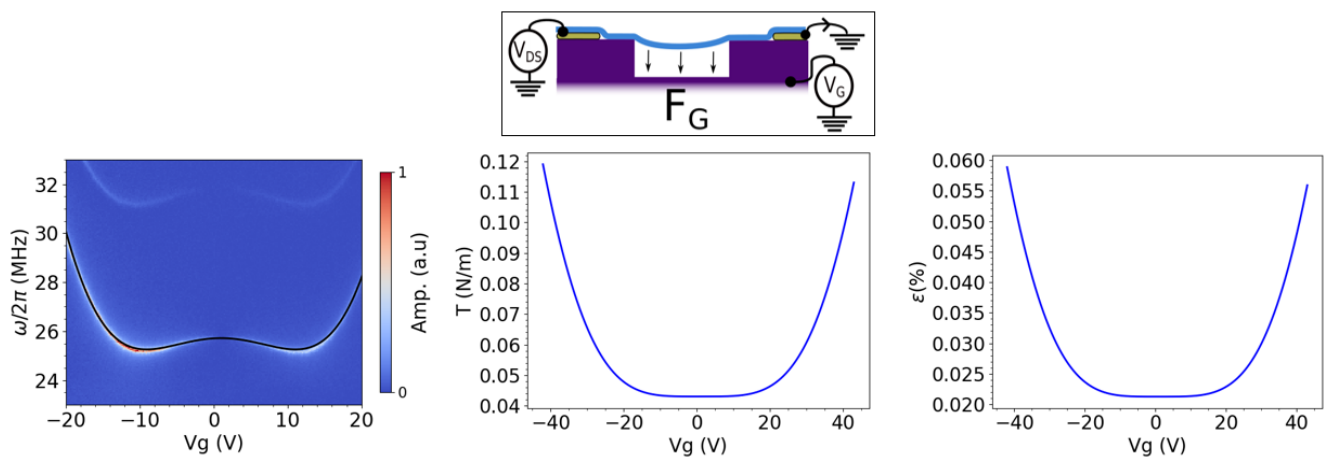

**Figure S7a: Optomechanic and static strain measurements in a 2D drum.** Measurement of the 2D drums vibration in function of the DC gate voltage following methods details in previous work [12,13]. The motion is actuated by an electrostatic force with a simple AC gate voltage applied to the substrate and measure with the confocal reflected laser (10µW at 633nm). The vibration vary with the DC gate voltage since the static force is in $V_G^2$. A softening at low $V_G$ appeared due to the capacitive softening effect. We plot the corresponding tension and strain applied along the sample, determined by our mechanical model which shift between 0.023 and 0.055%.

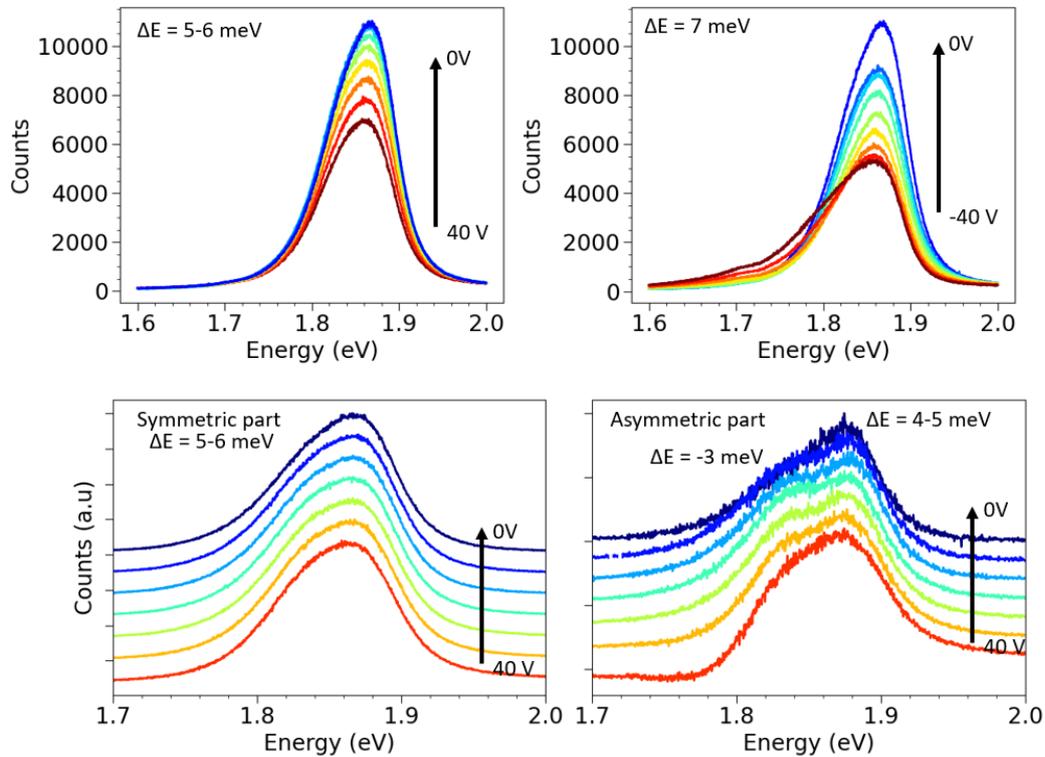

**Figure S7b: Photoluminescence measurements in function of the gate voltage.** We measured the PL spectrum at different gate voltage from -40V to 40V. Since the strain due to the electrostatic force is fully symmetric with the gate voltage (in $V_G^2$), we differentiate the PL signal response which is symmetric in $V_G$ and the PL signal which is asymmetric. The symmetric response shift by 5meV for 40V of gate voltage. Considering the strain difference of 0.04%, it gives a variation of 125meV/% of elongation along x, which corresponds to the simulated value.

## S8) Damaged membrane after measurements

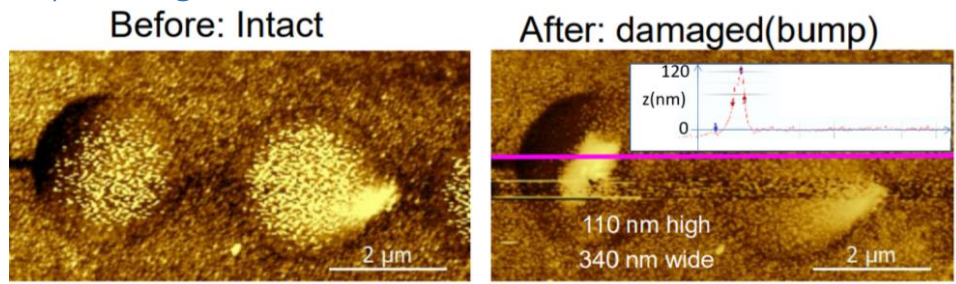

**Figure S8: Topography of the WSSe flake.** AFM image topography of the suspended flake before and after the last indentation applied to the sample. In this last measurement, only the far field measurement was recorded and we observe a small tent structure at the indentation point.

All the measurements in the publications are done before the breakdown of the membrane and we have confirmed by AFM topography that the membrane did not have any inelastic deformation after the maximum deflection of 250nm and 10% of strain.

## S9) Exciton Drift/Diffusion model in 2D- Auger rate and excitonic decay rate

The standard approach to steady-state bipolar transport in semiconductors is described by the Van Roosbroeck set of coupled differential equations.[14,15] It considers the Poisson's equation $-\nabla.\epsilon\nabla\varphi = q(C + p - n)$, the continuity equation for electrons $\nabla.\vec{J}_n = qR - qG$ and the continuity equation for holes $\nabla.\vec{J}_p = -qR + qG$. q is the charge q=1.6.10$^{-19}$C. p and n are the density of holes and electrons. $\epsilon$ is the dielectric permittivity. $\varphi$ is the electrostatic potential. C is the doping density. r and G are recombination rate and generation rate respectively. The electron and hole currents appearing in the continuity equations are $\vec{J}_n = -q\mu_n n\nabla\phi_n$ and $\vec{J}_p = -q\mu_p p\nabla\phi_p$. The independent variables in the problem are the potential $\varphi$, and the quasi-electrochemical potentials for electrons $\phi_n$ and holes $\phi_p$. The recombination terms R include a spontaneous radiative term, a Shockley-Read-Hall term [16] and an Auger term

$$R_{spont} = r_{spont}(np - n_i^2)$$
$$R_{SRH} = \frac{np - n_i^2}{(n + n_1)\tau_p + (p + p_1)\tau_n}$$
$$R_{Auger} = (np - n_i^2)(C_n n + C_p p)$$

$n_i$ is the initial density. $C_n$ and $p$ are coefficient for holes and electrons. The Van Roosbroeck equations are discretized using the Scharfetter-Gummel approach [17] which ensures current conservation. The strain dependence of the band structure is introduce trough the two electrochemical potentials, as is the case for the Seebeck effect. [18] Using non-degenerate electron statistics, this was implemented in Matlab to describe the exciton and trion diffusion in 2D material under strain. [16,19]

If we consider that the band gap is the only parameter modifying $\phi_n$ and $\phi_p$, we can simply reduce them to the band gap energy, $E_G$. In order to solve the drift-diffusion equation $\nabla(D\nabla n(r)) + \nabla(\mu\, n(r)\nabla E_G(r)) - n(r)/\tau - n^2(r)R_A + S(r) = 0$, we have also used the pdepe solver of matlab.

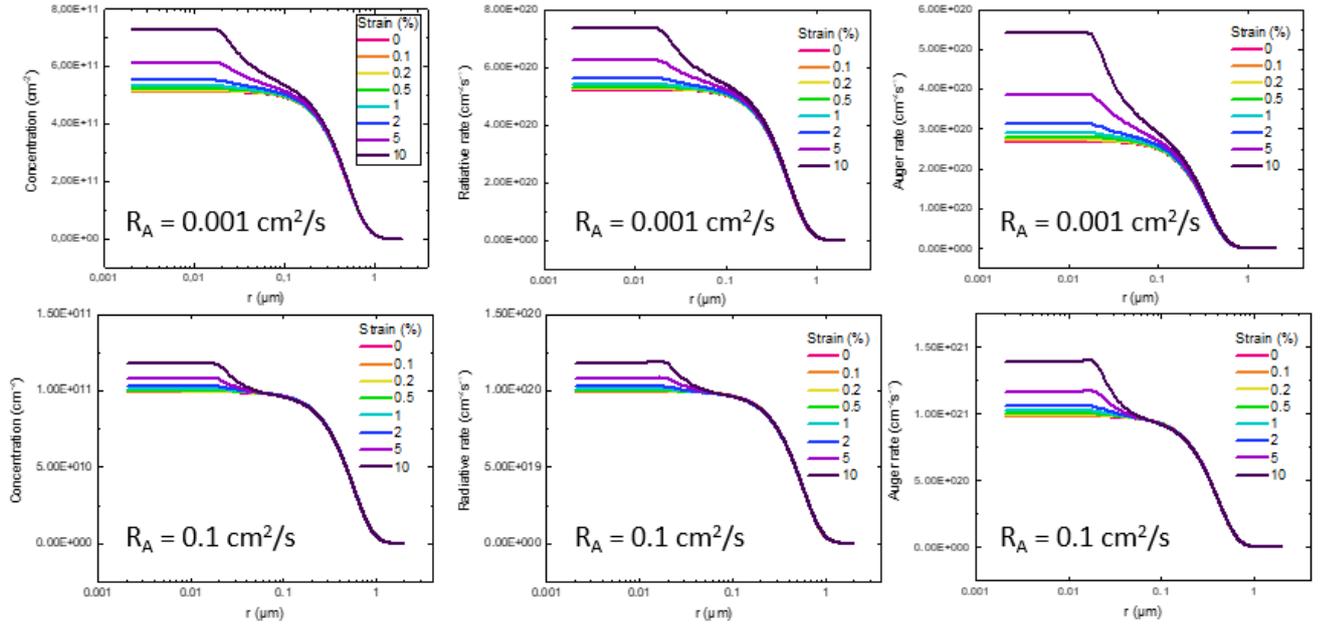

**Figure S9: Density of excitons in function of position, radial recombination, and auger recombination and strain applied to the membrane center.** We plot the concentration of exciton n, the radiative rate, the auger rate in function of the radius for different strain values from 0 to 10% and for two different values of $R_A$, the Auger coefficient. We note $\tau = $ n/radiative rate .

## S10) Comparison of the photoluminescence in contact mode or non-contact mode and for different 2D material ($WSe_2$, $WS_2$ and WSSe)

We did few tests to certify that the metallic tip does not affect the PL signal of the as-grown 2D material on substrate and of the suspended 2D membrane (noted transferred WSSe in the figure S10). We compare the PL signal when the tip is up and does not touch the 2D material and the PL signal when the tip is down (in contact with the 2D material with a very small force and deflection applied to the 2D). We did it for the as-grown material with two different laser and for the transferred WSSe. We can observe an increase of 22% of PL intensity for the as grown 2D with the 532 nm excitation, an increase of 46% for the as grown material with the 638 nm excitation and an increase of 12% for the suspended membrane under 532nm excitation. These increases of PL signal correspond to the TEPL signal. We do not observe any shift or difference in shape for any of the measured signal, between the up and down situation, which means that the metallic nature of the tip does not affect strongly the photoluminescence signal and as far as we know, it does not affect the band structure of the material.

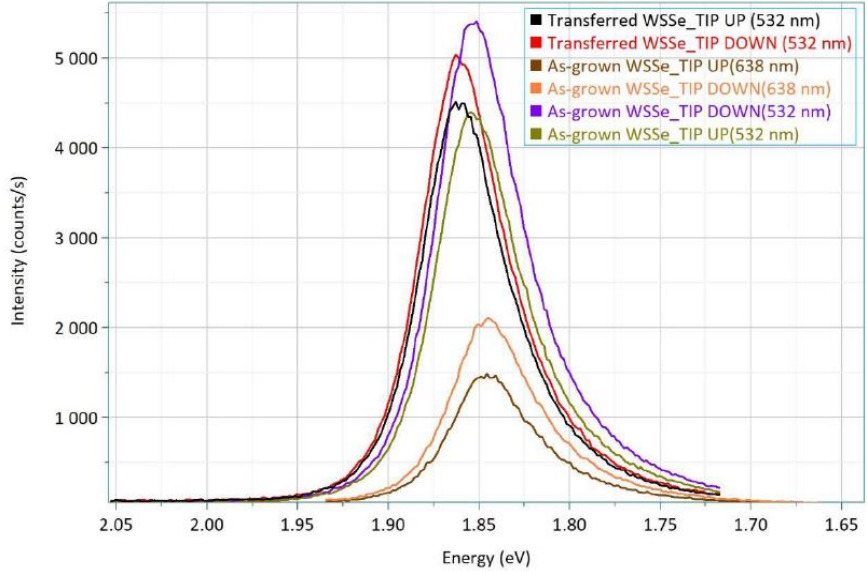

**Figure S10a: Difference of photoluminescence in contact or non-contact mode.** We plot and compare the PL spectrum when the tip is in contact with the 2D material (DOWN) and when the tip is not in contact with the 2D material (UP). We do it for three situations. The first situation is directly measured on the suspended 2D material. We only see an increase of intensity corresponding to the TEPL signal. The second and third test are references. They are done on as grown WSSe on the growth SOI substrate for two different lasers (532 nm and 638 nm).

To testify TEPL's accuracy, we have done measurements of pure $WS_2$ and $WSe_2$. Measurements were done in a lateral heterostructure made of $WSe_2$, WSSe and $WS_2$. The fabrication of such heterostructures has been already report by the authors.[20,21] A tip enhanced photoluminescence cartography has been used to selectively distinguished the three materials and to plot the different optical responses. We observe that the PL signal of the ternary is in between the PL signal position of the two binary and the amplitude is also very similar. This suggest that the ternary does not affect the TEPl measurement.

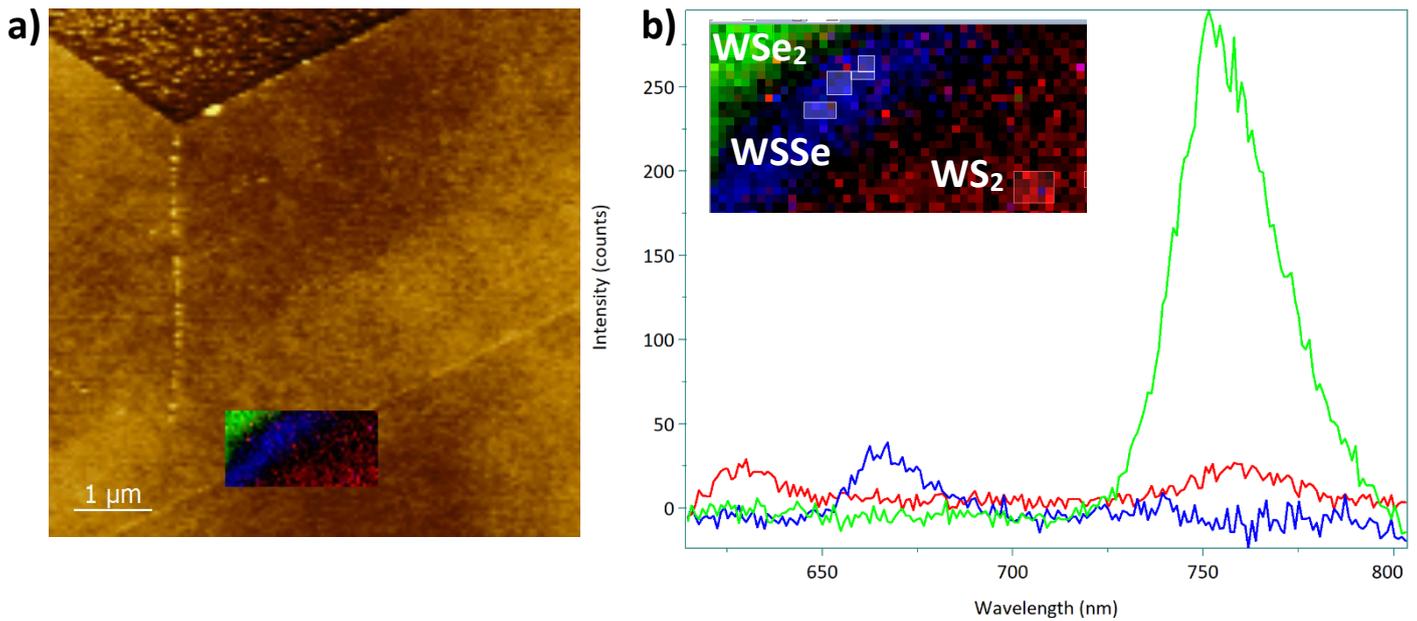

**Figure S10b: Tip enhanced Photoluminescence in a lateral heterostructure of $WS_2$, WSSe, $WSe_2$.** A) We measure the AFM image of a monolayer lateral heterostructure made of $WSe_2$, WSSe and $WS_2$. The WSSe is few hundreds of nm of width in this case. b) The TEPL spectrum with a 532nm illumination, for $WSe_2$(green), WSSe (blue) and $WS_2$ (red) In inset a cartography of the TEPL signal along the heterostructure with in red the intensity around 630nm, in blue the intensity around 670nm and in green the intensity around 750nm. The area is reported in the AFM image for clarity

**Bibliography**

[1]  E. Schwerin, *ZAMM - Journal of Applied Mathematics and Mechanics / Zeitschrift für Angewandte Mathematik und Mechanik* **1929**, *9*, 482.

[2]  C. Lee, X. Wei, J. W. Kysar, J. Hone, *Science* **2008**, *321*, 385.

[3]  S. Bertolazzi, J. Brivio, A. Kis, *ACS Nano* **2011**, *5*, 9703.

[4]  M. Annamalai, S. Mathew, M. Jamali, D. Zhan, M. Palaniapan, *J. Micromech. Microeng.* **2012**, *22*, 105024.

[5]  R. Zhang, V. Koutsos, R. Cheung, *Appl. Phys. Lett.* **2016**, *108*, 042104.

[6]  D. Vella, B. Davidovitch, *Soft Matter* **2017**, *13*, 2264.

[7]  J. Feng, X. Qian, C.-W. Huang, J. Li, *Nature Photon* **2012**, *6*, 866.

[8]  H. Moon, G. Grosso, C. Chakraborty, C. Peng, T. Taniguchi, K. Watanabe, D. Englund, *Nano Lett.* **2020**, *20*, 6791.

[9]  M. R. Begley, T. J. Mackin, *Journal of the Mechanics and Physics of Solids* **2004**, *52*, 2005.

[10] C. Jin, A. Davoodabadi, J. Li, Y. Wang, T. Singler, *Journal of the Mechanics and Physics of Solids* **2017**, *100*, 85.

[11] N. M. Bhatia, W. Nachbar, *AIAA Journal* **1968**, *6*, 1050.

[12] J. Chaste, I. Hnid, L. Khalil, C. Si, A. Durnez, X. Lafosse, M.-Q. Zhao, A. T. C. Johnson, S. Zhang, J. Bang, A. Ouerghi, *ACS Nano* **2020**, *14*, 13611.

[13] A. Chiout, F. Correia, M.-Q. Zhao, A. T. C. Johnson, D. Pierucci, F. Oehler, A. Ouerghi, J. Chaste, *Appl. Phys. Lett.* **2021**, *119*, 173102.

[14] W. van Roosbroeck, *Phys. Rev.* **1953**, *91*, 282.

[15] S. Park, B. Han, C. Boule, D. Paget, A. C. H. Rowe, F. Sirotti, T. Taniguchi, K. Watanabe, C. Robert, L. Lombez, B. Urbaszek, X. Marie, F. Cadiz, *2D Mater.* **2021**, *8*, 045014.

[16] H. Li, A. Thayil, C. T. K. Lew, M. Filoche, B. C. Johnson, J. C. McCallum, S. Arscott, A. C. H. Rowe, *Phys. Rev. Appl.* **2021**, *15*, 014046.

[17] D. L. Scharfetter, H. K. Gummel, *IEEE Transactions on Electron Devices* **1969**, *16*, 64.

[18] S. Park, Polarisation Resolved Photoluminescence in van Der Waals Heterostructures, These de doctorat, Institut polytechnique de Paris, **2022**.

[19] H. Li, Trap Mediated Piezoresponse of Silicon in the Space Charge Limit., phdthesis, Université Paris Saclay (COmUE), **2019**.

[20] B. Zheng, C. Ma, D. Li, J. Lan, Z. Zhang, X. Sun, W. Zheng, T. Yang, C. Zhu, G. Ouyang, G. Xu, X. Zhu, X. Wang, A. Pan, *J. Am. Chem. Soc.* **2018**, *140*, 11193.

[21] C. Ernandes, L. Khalil, H. Almabrouk, D. Pierucci, B. Zheng, J. Avila, P. Dudin, J. Chaste, F. Oehler, M. Pala, F. Bisti, T. Brulé, E. Lhuillier, A. Pan, A. Ouerghi, *npj 2D Materials and Applications* **2021**, *5*, 1.